\pdfoutput=1
 
\documentclass[cernpreprint,atlasdraft=false,texlive=2016,UKenglish,texmf]{atlasdoc}

\usepackage[subfig]{atlaspackage}

\usepackage{atlasbiblatex}
 
\usepackage{atlasphysics}
 
\graphicspath{{logos/}{figures/}}
 
\usepackage{ANA-HDBS-2019-22-PAPER-defs}
\usepackage{tabularx}

\AtlasTitle{Reconstruction and identification of boosted di-$\tau$ systems in a search for Higgs boson pairs using 13\,\TeV\ proton--proton collision data in ATLAS}

\AtlasAbstract{
In this paper, a new technique for reconstructing and identifying hadronically decaying $\tau^+\tau^-$
pairs with a large Lorentz boost, referred to as the di-$\tau$ tagger, is developed and used for the
first time in the ATLAS experiment at the Large Hadron Collider. A benchmark di-$\tau$ tagging selection
is employed in the search for resonant Higgs boson pair production, where one Higgs boson decays into a
boosted $b\bar{b}$ pair and the other into a boosted $\tau^+\tau^-$ pair, with two hadronically decaying
$\tau$-leptons in the final state. Using 139\,\ifb\ of proton--proton collision data recorded at a
centre-of-mass energy of 13\,\TeV, the efficiency of the di-$\tau$ tagger is determined and the background
with quark- or gluon-initiated jets misidentified as di-$\tau$ objects is estimated. The search for
a heavy, narrow, scalar resonance produced via gluon--gluon fusion and decaying into two Higgs bosons
is carried out in the mass range 1--3\,\TeV\ using the same dataset. No deviations from the Standard
Model predictions are observed, and 95\% confidence-level exclusion limits are set on this model.
}

\AtlasRefCode{HDBS-2019-22}
 
\PreprintIdNumber{CERN-EP-2020-118}

\arXivId{2007.14811}

\AtlasJournalRef{\JHEP 11 (2020) 163}
\AtlasDOI{https://doi.org/10.1007/JHEP11(2020)163}
 
\hypersetup{pdftitle={ATLAS document},pdfauthor={The ATLAS Collaboration}}
 
\begin{document}
 
\maketitle
 
\tableofcontents
 
\section{Introduction}
\label{sec:intro}
The discovery of the Higgs boson ($H$) by the ATLAS and CMS collaborations at the Large Hadron
Collider (LHC) in 2012~\cite{HIGG-2012-27,CMS-HIG-12-028} opens new ways of probing physics
beyond the Standard Model (SM), since the Higgs boson may itself appear as one of the intermediate
states in the decay of new resonances. Various final states have been used by ATLAS and CMS in searches
for both resonant and non-resonant $HH$ production~\cite{HIGG-2016-16,CMS-HIG-17-002,HDBS-2018-58,
CMS-HIG-17-030,HDBS-2018-33,HDBS-2018-18}. In the high-mass regime, for resonance masses typically
above 1\,\TeV, the Higgs bosons may be produced with large momenta, causing their decay products to be collimated.
The standard reconstruction techniques become inefficient in this regime. Therefore, a new technique,
referred to as the di-$\tau$ tagger, is developed to reconstruct and identify boosted hadronically decaying
$\tau^+\tau^-$ pairs. For the identification, a multivariate algorithm is trained to distinguish between
$\tau^+\tau^-$ pairs and the multi-jet background from quark- or gluon-initiated jets by exploiting the
calorimetric shower shapes and tracking information. A similar algorithm was implemented by the CMS Collaboration
in Ref.~\cite{CMS-TAU-16-003}.
 
An application of the di-$\tau$ tagger is carried out in a search for a narrow spin-0 resonance in the
mass range 1--3\,\TeV, which is produced via gluon--gluon fusion and decays into a pair of Higgs bosons
($X \to HH$), as predicted by models with an extended Higgs sector, such as two-Higgs-doublet
models~\cite{Lee:1973iz,Branco:2011iw}. This search considers the final state where one Higgs boson decays
into a $b\bar{b}$ pair and the other one into a $\tau^+\tau^-$ pair, where both $\tau$-leptons decay
hadronically.\footnote{SM branching fractions of the Higgs boson are assumed throughout the paper.}
A dedicated benchmark of the di-$\tau$ tagger with an identification efficiency of 60\% is designed for this
analysis. Using 139\,\ifb\ of proton--proton ($pp$) collision data at a centre-of-mass energy $\sqrt{s} = 13\,\TeV$
recorded by the ATLAS experiment in 2015--2018, various orthogonal event categories are defined in order to
correct the efficiency of the di-$\tau$ tagger for the benchmark selection, to perform and validate the multi-jet
background estimate, and to search for resonantly produced Higgs boson pairs.
 
This paper is organised as follows. After a brief description of the ATLAS detector in Section~\ref{sec:detector},
the samples of data and simulated events used in this study are described in Section~\ref{sec:samples}.
The procedures used to reconstruct and identify physics objects such as electrons, muons, jets and missing
transverse momentum in the detector are described in Section~\ref{sec:objects}. Section~\ref{sec:ditau} presents
the reconstruction and identification of boosted hadronically decaying $\tau^+\tau^-$ pairs.
In Section~\ref{sec:evtselection}, general event selections and categorisations are summarised, while
Section~\ref{sec:ddbkgest} focuses on the data-driven estimation of the multi-jet background with quark- or
gluon-initiated jets misidentified as boosted hadronically decaying $\tau^+\tau^-$ pairs. The data-driven
correction of the di-$\tau$ tagger efficiency is discussed in Section~\ref{sec:ditau-calib} and the search
for $X \to HH \to b\bar{b}\tau^+\tau^-$ is presented in Section~\ref{sec:hhbbtautau}, including the
statistical analysis used to set 95\% confidence-level (CL) limits on the production cross-section for
resonant $HH$ production. Finally, a summary is given in Section~\ref{sec:conclusion}.
 
\section{ATLAS detector}
\label{sec:detector}
The ATLAS detector~\cite{PERF-2007-01} at the LHC is a multipurpose particle detector
with a forward--backward symmetric cylindrical geometry and nearly \(4\pi\) coverage in
solid angle.\footnote{
ATLAS uses a right-handed coordinate system with its origin at the nominal interaction point (IP)
in the centre of the detector and the \(z\)-axis along the beam pipe.
The \(x\)-axis points from the IP to the centre of the LHC ring,
and the \(y\)-axis points upwards.
Cylindrical coordinates \((r,\phi)\) are used in the transverse plane,
\(\phi\) being the azimuthal angle around the \(z\)-axis.
The pseudorapidity is defined in terms of the polar angle \(\theta\) as \(\eta = -\ln \tan(\theta/2)\).
Angular distance is measured in units of \(\Delta R \equiv \sqrt{(\Delta\eta)^{2} + (\Delta\phi)^{2}}\).
}
It consists of an inner tracking system surrounded by a thin superconducting solenoid
providing a \SI{2}{\tesla} axial magnetic field, electromagnetic and hadronic calorimeters,
and a muon spectrometer.
 
The inner detector covers the pseudorapidity range \(|\eta| < 2.5\). It consists of silicon pixel, silicon microstrip,
and transition radiation tracking detectors. For the $\sqrt{s} = 13$~\TeV\ run, a fourth layer of the pixel detector,
the insertable $B$-layer~\cite{ATLAS-TDR-2010-19,PIX-2018-001}, was installed close to the beam pipe
at an average radius of 33.2~mm, providing an additional position measurement with $8~\mu$m resolution in the
$(x, y)$ plane and $40~\mu$m along $z$.
 
Lead/liquid-argon (LAr) sampling calorimeters provide electromagnetic energy measurements with high granularity
in the region $|\eta| < 3.2$. In the central part, $|\eta| < 2.5$, the calorimeter is divided into three layers,
one of them segmented in thin $\eta$ strips for optimal $\gamma/\pi_0$ separation, completed by a presampler layer
for $|\eta| < 1.8$. A hadronic steel/scintillator-tile calorimeter covers the central pseudorapidity range
(\(|\eta| < 1.7\)). The endcap and forward regions are instrumented with LAr calorimeters for both the
electromagnetic and hadronic energy measurements up to \(|\eta| = 4.9\). The granularity of the calorimeter
system in terms of $\Delta \eta \times \Delta \phi$ is typically $0.025 \times \pi/128$ in the barrel of the
electromagnetic calorimeter and $0.1 \times \pi/32$ in the hadronic calorimeter, with variations in segmentation
with $|\eta|$ and the layer~\cite{PERF-2014-07}.
 
The muon spectrometer surrounds the calorimeters and is based on three large air-core toroidal superconducting
magnets with eight coils each. The field integral of the toroids ranges between \num{2.0} and \SI{6.0}{\tesla\metre}
across most of the detector. The muon spectrometer includes a system of precision tracking chambers and fast
detectors for triggering.
 
A two-level trigger system~\cite{TRIG-2016-01} is used to select events. The first-level trigger is implemented
in hardware and uses a subset of the detector information to reduce the accepted rate to at most \SI{100}{\kHz}.
This is followed by a software-based trigger that reduces the accepted event rate to \SI{1}{\kHz} on average.
 
\section{Data and simulated events}
\label{sec:samples}
The studies presented in this paper are performed using a sample of $pp$ collision data recorded at
a centre-of-mass energy $\sqrt{s}= 13\,\TeV$ between 2015 and 2018, during stable beam conditions
and when all detector components relevant to the analysis were operating nominally~\cite{Aad:2019hmz}.
This corresponds to an integrated luminosity of 139\,\ifb. Samples of Monte Carlo (MC) simulated events
are used to train and calibrate the di-$\tau$ tagger, as well as to model the signal and some SM background
processes in the search for resonant Higgs boson pair production.
 
The signal, i.e.\ the production of a heavy spin-0 resonance via gluon--gluon fusion and its decay into a
pair of Higgs bosons, $X \to HH$, was simulated for nine values of the resonance mass, $m_X$,
between 1 and 3\,\TeV, using \mgamc~v2.6.1~\cite{Alwall:2014hca} at leading-order (LO) accuracy in quantum
chromodynamics (QCD) with the NNPDF2.3LO~\cite{Ball:2014uwa} set of parton distribution functions (PDFs). The
event generator was interfaced with \herwig~v7.1.3~\cite{Bahr:2008pv,Bellm:2015jjp} to model the parton shower,
hadronisation and underlying event, using the default set of tuned parameters (tune) and the
MMHT2014LO~\cite{Harland-Lang:2014zoa} PDF set. In the nine signal samples, a narrow-width approximation
was used for the resonance $X$, i.e.\ its natural width was set to a value that remains much smaller
than the experimental mass resolution. In addition, the Higgs boson mass was set to 125\,\GeV\ and the
SM branching fractions were used for the decays $H \to b\bar{b}$ and $H \to \tau^+\tau^-$.
In order to develop the identification algorithm of the boosted hadronically decaying $\tau^+\tau^-$
pairs, another set of $HH$ samples was produced, based on a narrow-width spin-2 Kaluza--Klein
graviton, as predicted in the Randall--Sundrum model of warped extra dimensions~\cite{Randall:1999ee},
$G \to HH \to (\tau^+\tau^-)(\tau^+\tau^-)$. Such events were generated with \mgamc~v2.3.3 at LO accuracy
in QCD with the NNPDF2.3LO set of PDFs, interfaced with \textsc{Pythia}~v8.212~\cite{Sjostrand:2014zea}
using the A14~\cite{ATL-PHYS-PUB-2014-021} tune. Five samples, with graviton masses of 1.5, 1.75, 2, 2.25 and
2.5\,\TeV, were generated. Only hadronically decaying $\tau$-leptons were considered in all $HH$ signal samples.
 
The production of $W$ and $Z$ bosons in association with jets ($V$+jets) was simulated with
\sherpa~v2.2.1~\cite{Bothmann:2019yzt} using matrix elements at next-to-leading-order (NLO)
accuracy in QCD for up to two jets and at LO accuracy for up to four jets, calculated with the
Comix~\cite{Gleisberg:2008fv} and OpenLoops~\cite{Cascioli:2011va} libraries. They were
matched with the \sherpa parton shower~\cite{Schumann:2007mg} using the MEPS@NLO
prescription~\cite{Hoeche:2009rj,Hoeche:2012yf}. The tune developed by the \sherpa authors
and the NNPDF3.0NNLO PDF set were used. The $V$+jets samples were normalised to a
next-to-next-to-leading-order (NNLO) prediction~\cite{Catani:2009sm}. In the $Z$+jets events,
jets are labelled according to the generated hadrons with $\pt > 5\,\GeV$ found within
a cone of size $\Delta R = 0.4$ around the jet axis. If a $b$-hadron is found, the jet is
labelled as a $b$-jet. If no $b$-hadron is found but there is a $c$-hadron instead, the jet is
labelled as a $c$-jet. If neither a $b$-hadron nor a $c$-hadron is found, the jet is labelled
as a light ($l$) jet. Simulated $Z$+jets events are then categorised according to the labels
of the two jets that are used to reconstruct the $H \to b\bar{b}$ candidate. The combination of
$Z$+$bb$, $Z$+$bc$, $Z$+$bl$ and $Z$+$cc$ events is referred to as $Z$+hf (denoting heavy-flavour
jets) in the following, whereas other events belong to the $Z$+lf (denoting light-flavour jets) category.
This categorisation is not performed for the $W$+jets process because its contribution is small.
 
The \POWHEGBOXV{v2} generator~\cite{Nason:2004rx,Frixione:2007vw,Alioli:2010xd} was used to generate
the $WW$, $WZ$ and $ZZ$ (diboson) processes~\cite{Nason:2013ydw} at NLO accuracy in QCD. The effect
of singly resonant amplitudes, as well as interference effects due to $Z/\gamma^*$ and identical
leptons in the final state, were included where appropriate (interference effects between $WW$ and
$ZZ$ for same-flavour charged leptons and neutrinos were ignored). Events were interfaced with
\textsc{Pythia}~v8.186~\cite{Sjostrand:2007gs} for the modelling of the parton shower, hadronisation
and underlying event, with parameters set according to the AZNLO~\cite{STDM-2012-23} tune. The
CT10~\cite{Lai:2010vv} PDF set was used for the hard-scattering processes, whereas the
CTEQ6L1~\cite{Pumplin:2002vw} PDF set was used for the parton shower.
 
The production of a single 125\,\GeV\ Higgs boson in association with a $Z$ boson was simulated up to
NLO accuracy in QCD using \POWHEGBOXV{v2}~\cite{Luisoni:2013kna, Cullen:2011ac, Hamilton:2012np}, with
the NNPDF3.0NLO PDF set and subsequently reweighted to the PDF4LHC15NLO~\cite{Butterworth:2015oua}
PDF set. The simulation was interfaced with \textsc{Pythia}~v8.212, using the AZNLO tune and the
CTEQ6L1 PDF set. The $gg \to ZH$ and $qq \to ZH$ samples were normalised to cross-sections calculated
at, respectively, NLO accuracy in QCD including soft-gluon resummation up to next-to-leading
logarithms~\cite{Altenkamp:2012sx,Harlander:2014wda,Hespel:2015zea} and NNLO accuracy in QCD
with NLO electroweak corrections~\cite{Brein:2012ne,Harlander:2018yio,Brein:2003wg,Brein:2011vx,Ferrera:2017zex,Ciccolini:2003jy,Denner:2011id,Denner:2014cla}.
Other single-Higgs-boson production modes were found to contribute negligibly.
 
Single-top-quark processes (split into $s$-channel, $t$-channel and $tW$ contributions) and \ttbar\ events
were simulated using \POWHEGBOXV{v2}~\cite{Frixione:2007nw,Alioli:2009je,Re:2010bp} at NLO accuracy in QCD with
the NNPDF3.0NLO PDF set. All events were interfaced with \textsc{Pythia}~v8.230 using the A14 tune and the
NNPDF2.3LO PDF set. For the $tW$ process, the diagram removal scheme~\cite{Frixione:2008yi} was employed
in order to handle the interference with \ttbar\ production. The \ttbar\ sample was normalised to the
cross-section prediction at NNLO accuracy in QCD including the resummation of next-to-next-to-leading
logarithmic soft-gluon terms calculated using \textsc{Top++2.0}~\cite{Beneke:2011mq,Cacciari:2011hy,Baernreuther:2012ws,Czakon:2012zr,Czakon:2012pz,Czakon:2013goa,Czakon:2011xx}.
For the single-top-quark processes, the cross-sections of the $s$- and $t$-channels were corrected to the theory
prediction at NLO accuracy in QCD calculated with \textsc{Hathor}~v2.1~\cite{Aliev:2010zk,Kant:2014oha}, while
the cross-section used for the $tW$ sample was based on the approximate NNLO calculation~\cite{Kidonakis:2010ux,Kidonakis:2013zqa}.
 
Except when using \sherpa, $b$- and $c$-hadron decays were performed with \evtgen~v1.2.0 or~v1.6.0~\cite{EvtGen},
while the decays of $\tau$-leptons were handled internally by all event generators. The effect of
multiple interactions in the same and neighbouring bunch crossings (pile-up) was modelled by overlaying
the original hard-scattering event with simulated inelastic $pp$ events generated with \textsc{Pythia}~v8.186 using the
NNPDF2.3LO PDF set and the A3~\cite{ATL-PHYS-PUB-2016-017} tune. The MC samples were
processed with a simulation~\cite{SOFT-2010-01} of the detector response based on
\textsc{Geant4}~\cite{Agostinelli:2002hh} and events were then reconstructed with
the same software as the data.
 
\section{Object reconstruction}
\label{sec:objects}
The following procedures are used to reconstruct and identify objects, such as electrons, muons,
jets and missing transverse momentum, in the ATLAS experiment.
 
In general, jets refer to the hadronic objects reconstructed using the \antikt\ algorithm with a radius
parameter $R=0.4$~\cite{Cacciari:2008gp,Fastjet}, starting from topological clusters of energy deposits in
the calorimeter. When used in the following, jets are required to have $\pT > 20\,\GeV$ and $|\eta|<4.5$. The
reconstruction of electrons is based on matching inner-detector tracks to energy clusters in the electromagnetic
calorimeter. Electrons are required to have $\pT > 7\,\GeV$ and $|\eta|<2.47$, excluding the barrel--endcap
transition region of the calorimeter ($1.37 < |\eta| < 1.52$). They are then identified using the `loose'
operating point provided by a likelihood-based algorithm~\cite{EGAM-2018-01}. The reconstruction of muons
relies on matching tracks in the inner detector and the muon spectrometer. Muons are required to have
$\pT > 7\,\GeV$ and $|\eta|<2.5$, as well as to satisfy the `loose' identification criteria and
`FixedCutLoose' isolation working point defined in Ref.~\cite{PERF-2015-10}.
 
In order to avoid double-counting of objects that overlap geometrically, an electron is removed if
it shares an inner-detector track with a muon. Then, \antikt\ jets with $R=0.4$ are discarded if they
meet one of the following two conditions: (i) $\Delta R(\text{jet}, e)<0.2$; (ii) the jet has less
than three associated tracks and either a muon inner-detector track is associated with the jet or
$\Delta R(\text{jet}, \mu)<0.2$. Finally, an electron or muon is discarded if found within a distance
$\Delta R = \mathrm{min}(0.4,\,0.04+10\,\GeV/\pT^{e/\mu})$ from any remaining jet.
 
The missing transverse momentum, the magnitude of which is denoted by \met in the following, is
defined as the negative vector sum of the transverse momenta of all fully reconstructed and calibrated
objects, after an overlap removal procedure that is distinct from that used for the electron/muon/jet
disambiguation above~\cite{PERF-2016-07,ATLAS-CONF-2018-023}. The missing transverse momentum also includes
a soft term, calculated using the inner-detector tracks that originate from the primary vertex
(defined as that having the largest sum of squared track-\pT) but are not associated with reconstructed
objects.
 
In order to capture the decay products of boosted particles, such as Higgs bosons, another type
of jets, called large-radius jets, is employed. These are also formed using the \antikt\
algorithm, but with a radius parameter $R=1.0$ and are built from topological clusters of energy deposits
calibrated using the local hadronic cell weighting scheme~\cite{PERF-2014-07}. In the following,
the large-radius jet matched to the boosted $H \to b\bar{b}$ candidate is referred to as a `large-$R$
jet' while that of the boosted hadronically decaying $\tau^+\tau^-$ pair is referred to as a `di-$\tau$
object'.
 
While the reconstruction and identification of boosted $H \to \tau^+\tau^-$ decays employ new techniques
described in Section~\ref{sec:ditau}, a standard procedure is used for boosted $H \to b\bar{b}$ decays.
Large-$R$ jets are trimmed~\cite{Krohn:2009th} to remove the effects of pile-up and the underlying event.
Trimming proceeds by reclustering the original constituents of a large-$R$ jet into a collection of
$R=0.2$ sub-jets using the $k_t$ algorithm~\cite{Ellis:1993tq,Catani:1993hr} and removing any sub-jets
with $p_\text{T}^\text{sub-jet}/p_\text{T}^{J_0}<0.05$, where $p_\text{T}^\text{sub-jet}$ is the transverse
momentum of the sub-jet under consideration and $p_\text{T}^{J_0}$ that of the original (untrimmed) large-$R$
jet. The energy and mass scales of the trimmed jets are then calibrated using $p_\text{T}$- and $\eta$-dependent
calibration factors derived from simulation~\cite{PERF-2012-02}. After trimming, the large-$R$ jet is required to
have $\pT > 300\,\GeV$ and its mass is calculated using the combined mass technique with tracking and calorimeter
information as input~\cite{ATLAS-CONF-2016-035}.
 
In order to identify $b$-hadrons within a large-$R$ jet, variable-radius track-jets~\cite{Krohn:2009zg}
are reconstructed using the \antikt\ algorithm from inner-detector tracks with a jet-\pT\ dependent radius
parameter $R(p_\text{T})=\rho/p_\text{T}$, where $\rho$ determines how fast the effective size of the jet
decreases with its transverse momentum. The lower ($R_{\text{min}}$) and upper ($R_{\text{max}}$) cut-offs
prevent the jet from becoming too large at low \pT\ and from shrinking below the detector resolution at
high \pT, respectively. In this paper, $\rho=30\,\GeV$, $R_{\text{min}}=0.02$ and $R_{\text{max}}=0.4$
are used~\cite{ATL-PHYS-PUB-2017-010}.
These track-jets are then matched to an untrimmed large-$R$ jet by using the ghost-association
method~\cite{Cacciari:2008gn, Cacciari:2007fd}. Only track-jets with $\pT > 10\,\GeV$, $|\eta|<2.5$
and at least two tracks~\cite{ATL-PHYS-PUB-2017-010} are considered in the following. Events with two
collinear track-jets $a$ and $b$ that fulfil
$\Delta R (\text{jet}_a,\text{jet}_b) < \mathrm{min}(R_{\text{jet}_a}, R_{\text{jet}_b})$ are removed.
 
The flavour of track-jets is determined using a multivariate approach based on the properties and vertex
information of the associated tracks~\cite{PERF-2012-04, ATL-PHYS-PUB-2015-022}. Various $b$-jet identification
algorithms are used to exploit impact-parameter information, secondary-vertex information, and $b$- to $c$-hadron
decay chain information. The \textsc{MV2c10} algorithm~\cite{ATL-PHYS-PUB-2016-012} then combines information
from the various upstream algorithms in a boosted decision tree (BDT) that is trained to discriminate $b$-jets from
a background sample made of 93\% light-flavour jets and 7\% $c$-jets. In order to label a track-jet as $b$-tagged,
a requirement is placed on the output score of the \textsc{MV2c10} discriminant. This requirement has an average
efficiency of 70\% for $b$-jets in simulated $t\bar{t}$ events (decreasing to about 60\% for a $b$-jet $\pt$ above
500\,\GeV) with rejection factors of 8.9, 36 and 300 for jets initiated by $c$-quarks, hadronically decaying
$\tau$-leptons and light-flavour quarks, respectively~\cite{FTAG-2018-01}. The number of $b$-tagged track-jets
in the large-$R$ jet is used to define various event categories, as described in Section~\ref{sec:evtselection}.
 
\section{Reconstruction and identification of boosted hadronically decaying $\tau^+\tau^-$ pairs}
\label{sec:ditau}
Hadronically decaying $\tau$-leptons produce a neutrino and visible decay products, typically one or three
charged pions and up to several neutral pions, which are reconstructed and identified as \tauhadvis\ objects.
In the standard procedure~\cite{ATLAS-CONF-2017-029}, a \tauhadvis\ object is seeded by a jet with
$\pT > 10\,\GeV$ and $|\eta| < 2.5$, formed using the \antikt\ algorithm with $R=0.4$. A multivariate
identification stage, using calorimetric shower shapes and tracking information as input variables, is
employed to discriminate \tauhadvis\ candidates with one or three associated tracks from quark- or
gluon-initiated jets. However, in the search for high-mass $X \to HH \to b\bar{b}\tau^+\tau^-$ presented
in this paper, more than 50\% of the $\tau^+\tau^-$ pairs have a separation  $\Delta R < 0.4$ when
$m_X \ge 2\,\TeV$, hence they would fail the standard reconstruction procedure. For such events, a new
method for reconstructing boosted hadronically decaying $\tau^+\tau^-$ pairs, referred to as the
di-$\tau$ tagger, is employed.
 
Boosted di-$\tau$ objects are seeded by untrimmed large-radius jets that must have $\pT > 300\,\GeV$.
Their constituents are reclustered into \antikt\ sub-jets with $R=0.2$. The original di-$\tau$ seeding jet
must include at least two such sub-jets and, after ordering in \pT, the two leading
sub-jets are used to construct the di-$\tau$ system. Tracks are geometrically matched to a sub-jet if they
are within a cone of size $\Delta R = 0.2$ around its axis, and they are labelled as `$\tau$ tracks'. Other tracks
found in the isolation region (i.e.\ the area of the di-$\tau$ seeding large-radius jet excluding
the di-$\tau$ sub-jets) are labelled as `iso-tracks'. A schematic representation of a di-$\tau$
object is shown in Figure~\ref{fig:ditau-cartoon}.
The track selection criteria, as well as the track--vertex matching, are the same as those used for
the standard \tauhadvis\ objects~\cite{ATLAS-CONF-2017-029}. In the following, the two leading sub-jets
used to compute the four-momentum of the di-$\tau$ system must have $\pT > 10\,\GeV$ and at least one
associated track.
 
\begin{figure}[htbp]
\centering
\includegraphics[width=0.48\textwidth]{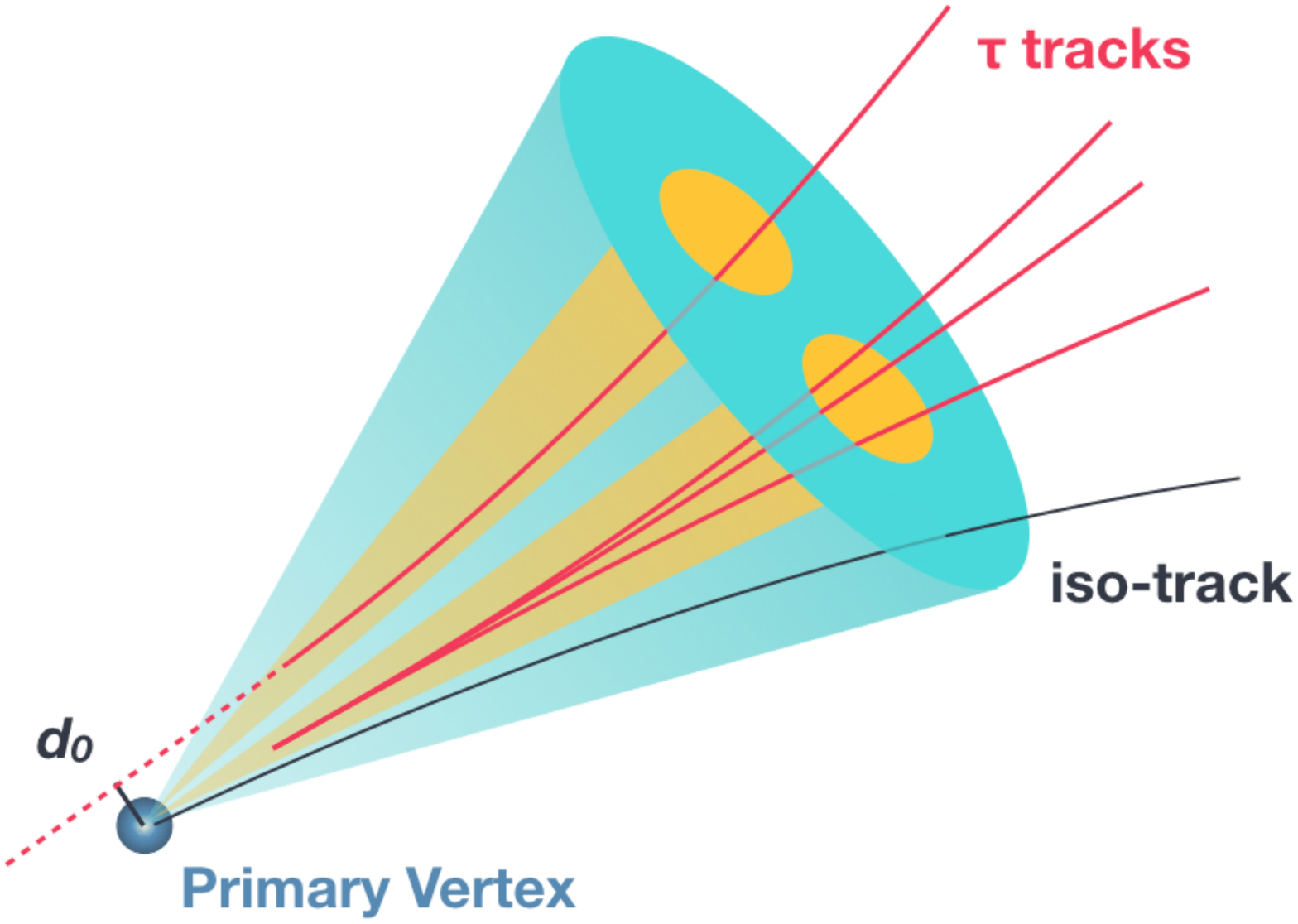}
\caption{Schematic representation of a di-$\tau$ object: the large blue cone is the di-$\tau$ seeding jet
while the two smaller yellow cones are the two leading sub-jets. Tracks found within a cone of size $\Delta R = 0.2$
around the sub-jet axis are matched to charged pions produced in hadronic decays of $\tau$-leptons,
while other tracks found in the isolation region are labelled as `iso-tracks'. The closest distance $d_0$
in the transverse plane between the primary vertex and the leading track matched to a sub-jet is also
shown for illustration.}
\label{fig:ditau-cartoon}
\end{figure}
 
At this stage, the di-$\tau$ reconstruction efficiency is defined as the fraction of events in which
a boosted di-$\tau$ candidate is reconstructed and each of the two leading sub-jets geometrically
matches a generated hadronically decaying $\tau$-lepton for which the \pT\ of the visible products
(neutral and charged hadrons) exceeds 10\,\GeV. Figures~\ref{fig:receffvsdr} and~\ref{fig:receffvspt}
show the di-$\tau$ reconstruction efficiency as a function of, respectively, the distance
$\Delta R(\tau_{1,\text{vis}},\tau_{2,\text{vis}})$ between the visible
products of the two hadronically decaying $\tau$-leptons and the visible
\pT\ of the di-$\tau$ system, both computed at generator level. For this
purpose, a sample of simulated $X \to HH \to b\bar{b}\tau^+\tau^-$ events is used, in
which the resonance mass is set to 2\,\TeV\ and both $\tau$-leptons decay hadronically. In
Figure~\ref{fig:rec-eff}, the reconstruction efficiency is also shown for standard (resolved)
\tauhadvis\ objects. In this case, the efficiency is defined as the fraction of events with at
least two reconstructed \tauhadvis\ candidates, with at least one associated track each, that
geometrically match a generated hadronically decaying $\tau$-lepton for which the \pT\ of
the visible products exceeds 10\,\GeV. This comparison shows that the boosted
di-$\tau$ object reconstruction method is necessary at high transverse momenta and
low $\Delta R(\tau_{1,\text{vis}},\tau_{2,\text{vis}})$.
In particular, Figure~\ref{fig:receffvsdr} shows that the reconstruction efficiencies decrease sharply
when the visible products of the two hadronically decaying $\tau$-leptons are so close that
they merge into one jet. With the resolved \tauhadvis\ reconstruction this happens when
$\Delta R(\tau_{1,\text{vis}},\tau_{2,\text{vis}})<0.4$, while the boosted di-$\tau$ reconstruction
extends the sensitivity down to $\Delta R(\tau_{1,\text{vis}},\tau_{2,\text{vis}})=0.2$ by resolving the
smaller sub-jets. In addition, as the distance $\Delta R(\tau_{1,\text{vis}},\tau_{2,\text{vis}})$ increases,
the reconstruction efficiency of both boosted di-$\tau$ and resolved \tauhadvis\ objects
decreases slowly, because the sub-leading generated $\tau$-lepton is found to become softer,
and hence less likely to exceed the 10\,\GeV\ \pT\ threshold imposed on its visible products.
Also, in contrast to the reconstruction efficiency of resolved \tauhadvis\ objects, which
is based on at least two candidates, the di-$\tau$ reconstruction method loses some efficiency
due to the fact that only the two leading sub-jets are considered.
Finally, as shown in Figure~\ref{fig:receffvspt}, the reconstruction efficiency reaches a plateau when
the \pT\ of the di-$\tau$ system exceeds 300\,\GeV, while the location of the turn-on is set by
the \pT\ cut on the seeding jet.

\begin{figure}[htbp]
\centering
\subfloat[]{
\includegraphics[width=0.46\textwidth]{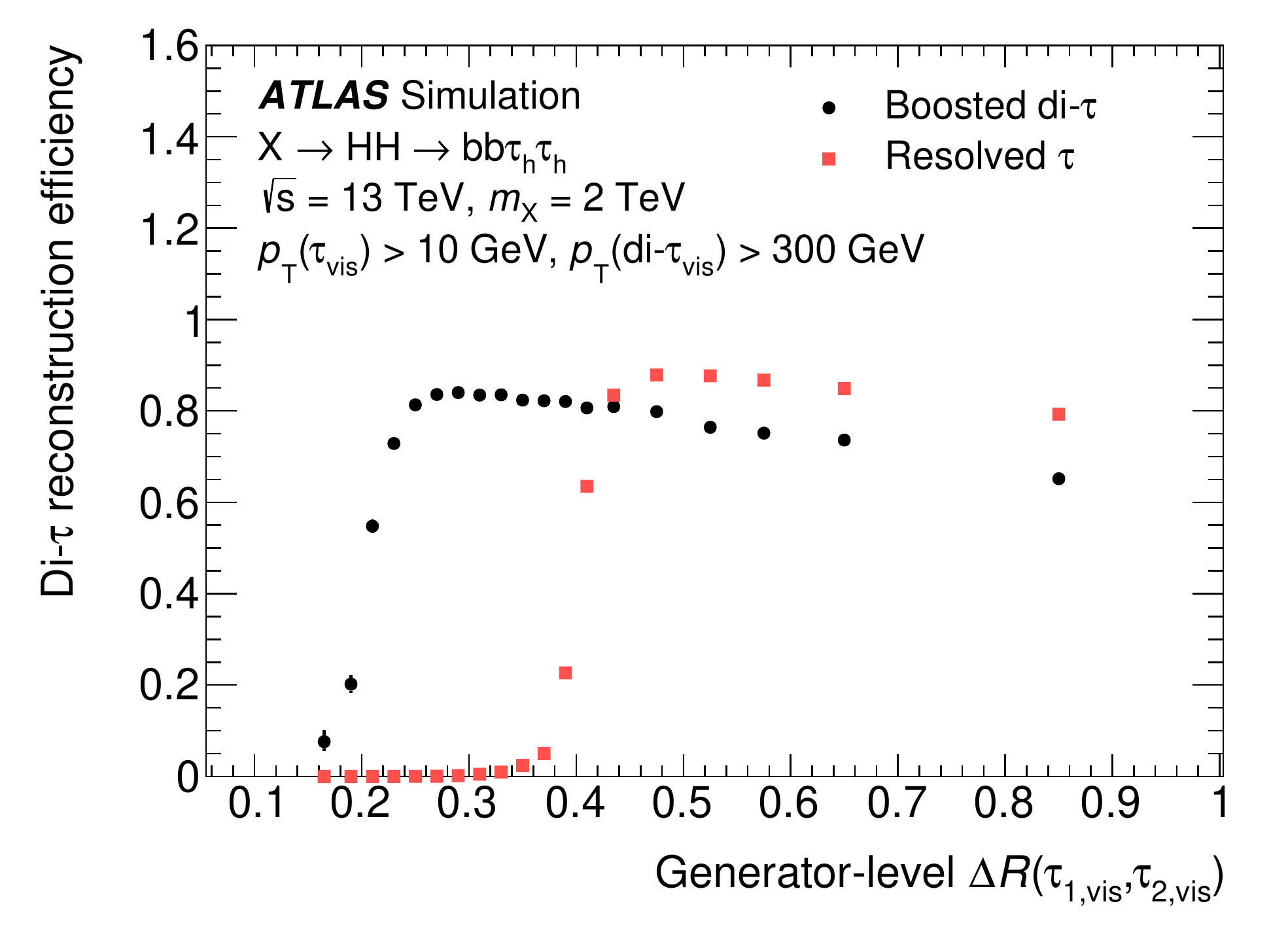}\label{fig:receffvsdr}
 
}
\subfloat[]{
\includegraphics[width=0.46\textwidth]{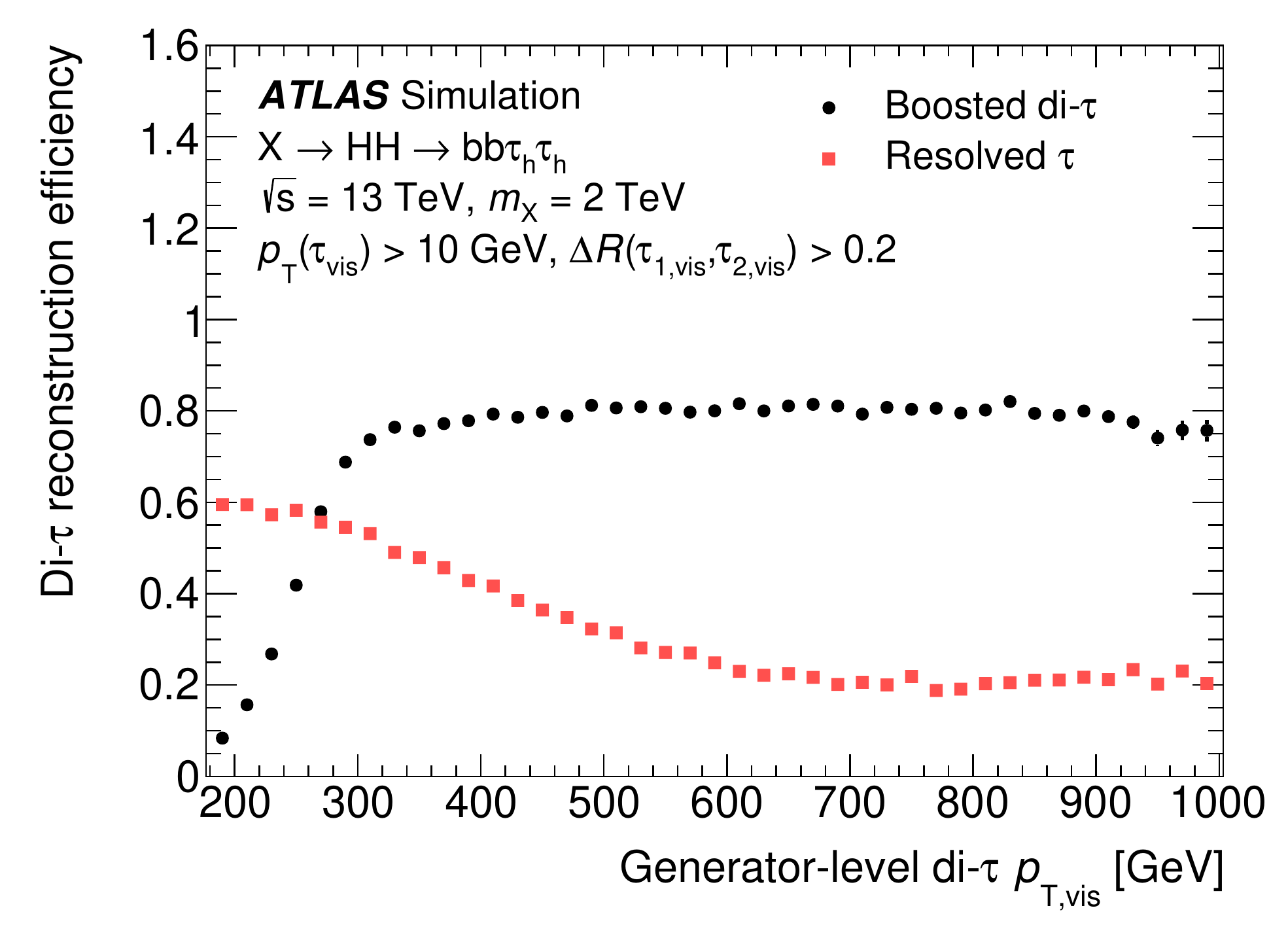}\label{fig:receffvspt}
}
\caption{Efficiency to reconstruct a di-$\tau$ system with (squares) resolved \tauhadvis\ objects and
(circles) a boosted di-$\tau$ object versus (a) the distance $\Delta R(\tau_{1,\text{vis}},\tau_{2,\text{vis}})$
between the visible products of the two hadronically decaying $\tau$-leptons and (b) the \pT\ of the
di-$\tau$ system, both at generator level. The reconstruction efficiency is computed in simulated
$X \to HH \to b\bar{b}\tau^+\tau^-$ events, where the resonance mass is set to 2\,\TeV\ and both
$\tau$-leptons decay hadronically (similar patterns are observed for other masses though).
The vertical error bars only account for statistical uncertainties.
}
\label{fig:rec-eff}
\end{figure}
 
As in the case of standard \tauhadvis\ objects, a separate identification stage using multivariate
techniques is employed to reduce the background from quark- and gluon-initiated jets. For this purpose,
a BDT discriminant is built using information about the clusters in the calorimeter, tracks and vertices. Multi-jet
events with quark- or gluon-initiated jets misidentified as di-$\tau$ objects are expected to have lower-$\pT$
sub-jets, with a larger fraction of energy in the isolation region. Also, multi-jet events typically have fewer
collimated tracks and a higher track multiplicity with, accordingly, a smaller fraction of the transverse momentum
carried by each track. The variables defined in Table~\ref{tab:idvars} are found to provide good discrimination
and are used in the BDT training.
 
\begin{table}[htbp]
\centering
\caption{Discriminating variables used in the di-$\tau$ identification BDT, aimed at rejecting the background
from quark- and gluon-initiated jets. Here, LRJ refers to the seeding large-radius jet of the di-$\tau$ object,
sj$_1$ and sj$_2$ stand for the first and second sub-jets ordered in \pT, respectively, and tracks refer to those
matched to a sub-jet ($\tau$ tracks), unless specified otherwise.}
\label{tab:idvars}
\small
\renewcommand{\arraystretch}{1.3}
\begin{tabularx}{\textwidth}{cX}
\toprule
Variable & Definition \\
\midrule
 
$E_{\Delta R < 0.1}^{\text{sj}_1}/E_{\Delta R < 0.2}^{\text{sj}_1}$ and $E_{\Delta R < 0.1}^{\text{sj}_2}/E_{\Delta R < 0.2}^{\text{sj}_2}$ & Ratios of the energy deposited in the core to that in the full cone, for the sub-jets sj$_1$ and sj$_2$, respectively \\
 
$\pT^{\text{sj}_2}/\pT^{\text{LRJ}}$ and $(\pT^{\text{sj}_1} + \pT^{\text{sj}_2})/\pT^{\text{LRJ}}$ & Ratio of the \pT\ of sj$_2$ to the di-$\tau$ seeding large-radius jet \pT\ and ratio of the scalar \pT\ sum of the two leading sub-jets to the di-$\tau$ seeding large-radius jet \pT, respectively \\
 
$\log(\sum\pT^\text{iso-tracks}/\pT^{\text{LRJ}})$ & Logarithm of the ratio of the scalar \pT\ sum of the iso-tracks to the di-$\tau$ seeding large-radius jet \pT\ \\
 
$\Delta R_\text{max}(\text{track,\,sj}_1)$ and $\Delta R_\text{max}(\text{track,\,sj}_2)$ & Largest separation of a track from its associated sub-jet axis, for the sub-jets sj$_1$ and sj$_2$, respectively\\
 
$\sum [\pT^\text{track}\Delta R(\text{track,\,sj}_2)] / \sum \pT^\text{track}$ & \pT-weighted $\Delta R$ of the tracks matched to sj$_2$ with respect to its axis \\
 
$\sum [\pT^\text{iso-track}\Delta R(\text{iso-track,\,sj})] / \sum \pT^\text{iso-track}$ & \pT-weighted sum of $\Delta R$ between iso-tracks and the nearest sub-jet axis \\
 
$\log(m_{\Delta R < 0.1}^{\text{tracks,\,sj}_1})$ and $\log(m_{\Delta R < 0.1}^{\text{tracks,\,sj}_2})$ & Logarithms of the invariant mass of the tracks in the core of sj$_1$ and sj$_2$, respectively \\
 
$\log(m_{\Delta R < 0.2}^{\text{tracks,\,sj}_1})$ and $\log(m_{\Delta R < 0.2}^{\text{tracks,\,sj}_2})$ & Logarithms of the invariant mass of the tracks with $\Delta R < 0.2$ from the axis of sj$_1$ and sj$_2$, respectively \\
 
$\log(|d_{0,\text{lead-track}}^{\text{sj}_1}|)$ and $\log(|d_{0,\text{lead-track}}^{\text{sj}_2}|)$ & Logarithms of the closest distance in the transverse plane between the primary vertex and the leading track of sj$_1$ and sj$_2$, respectively \\
 
$n_\text{tracks}^{\text{sj}_1}$ and $n_\text{tracks}^\text{sub-jets}$  & Number of tracks matched to sj$_1$ and to all sub-jets, respectively \\
\bottomrule
\end{tabularx}
\end{table}
 
The training of the BDT is performed using the adaptive boosting (AdaBoost) algorithm~\cite{Freund:1997xna},
with a boosting parameter of 0.5, in the TMVA toolkit~\cite{TMVA}. The di-$\tau$ objects in simulated events
with a spin-2 graviton, $G \to HH \to (\tau^+\tau^-)(\tau^+\tau^-)$, form the signal training set.
Five samples, with graviton masses of 1.5, 1.75, 2, 2.25 and 2.5\,\TeV, are combined to form the signal. The
spin of the resonance is found to have no impact on the di-$\tau$ identification.
 
The background sample in the BDT training consists of the 3.2\,\ifb\ of data recorded by the ATLAS experiment
in 2015 using combined jet triggers with transverse energy (\ET) thresholds between 100 and 400\,\GeV. It is
dominated by multi-jet events and the contamination from events with a true di-$\tau$ object is found to
be negligible. The di-$\tau$ \pT\ spectra of the signal and background events are reweighted to become flat,
so that the \pT\ dependencies of the input variables are eliminated in the BDT training. On the other hand,
no reweighting regarding $\eta$ is performed, since no dependency on this variable is observed.
Figure~\ref{fig:roc} shows the background rejection factor (defined as the inverse of the background
efficiency) versus the signal efficiency, as obtained after the di-$\tau$ reconstruction step. In
the following, a requirement is placed on the BDT output score to define a benchmark
with a signal efficiency of about 60\%, as indicated by the red cross. This working point
corresponds to a background rejection factor of $10^4$. Figure~\ref{fig:ideffvspt} shows that
the identification efficiency has little dependency on the \pT\ of the di-$\tau$ system
beyond 300\,\GeV. As illustrated in Figure~\ref{fig:ideffvsmu}, the performance of the BDT
is not affected much by the number of pile-up interactions. Hence, although the training
is performed using only data collected in 2015, the discrimination between the di-$\tau$ signature and
the multi-jet background is not expected to change significantly across the data-taking period between
2015 and 2018.
 
\begin{figure}[htbp]
\centering
\subfloat[]{
\includegraphics[width=0.45\textwidth]{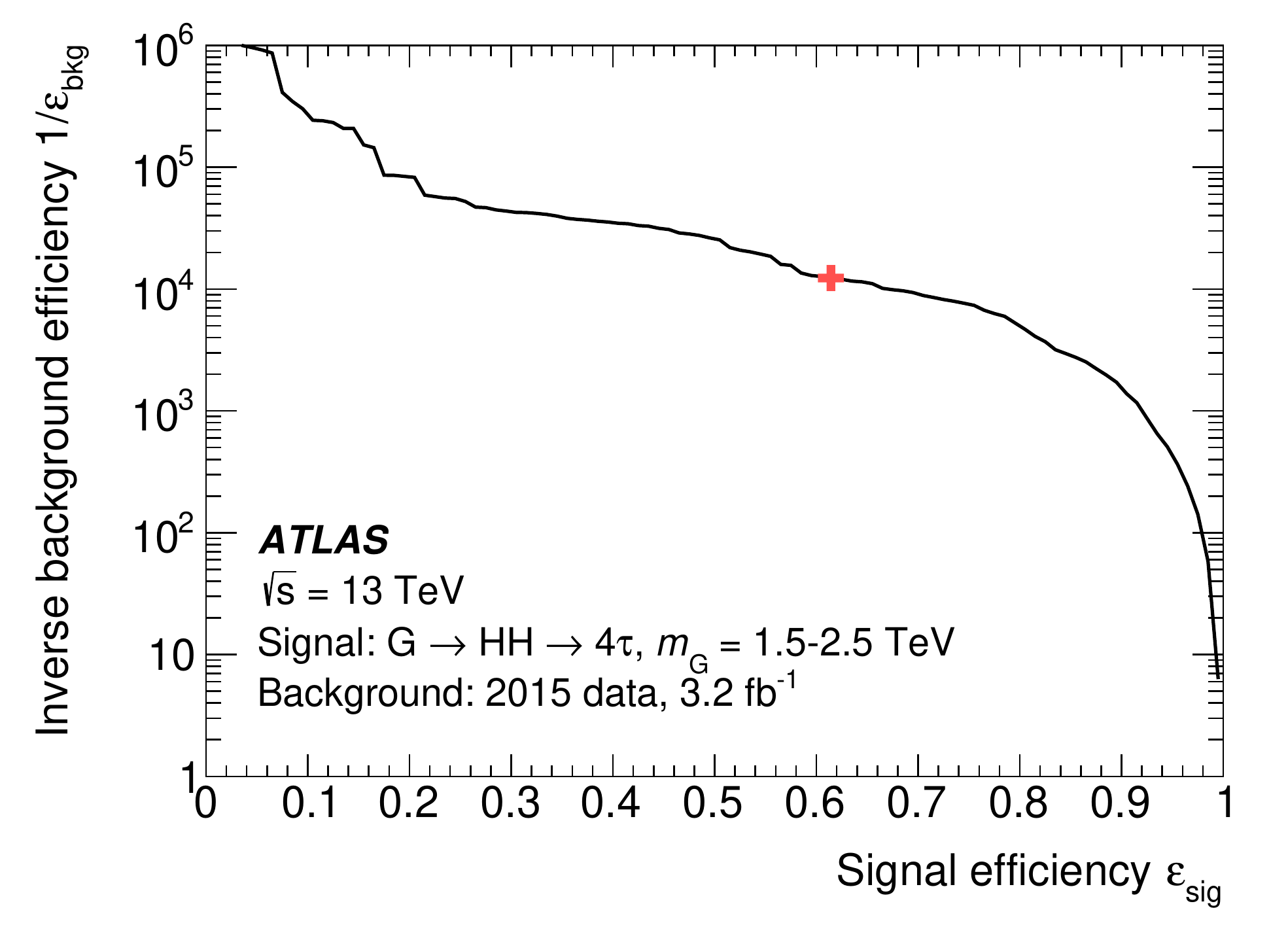}\label{fig:roc}
}\\
\subfloat[]{
\includegraphics[width=0.45\textwidth]{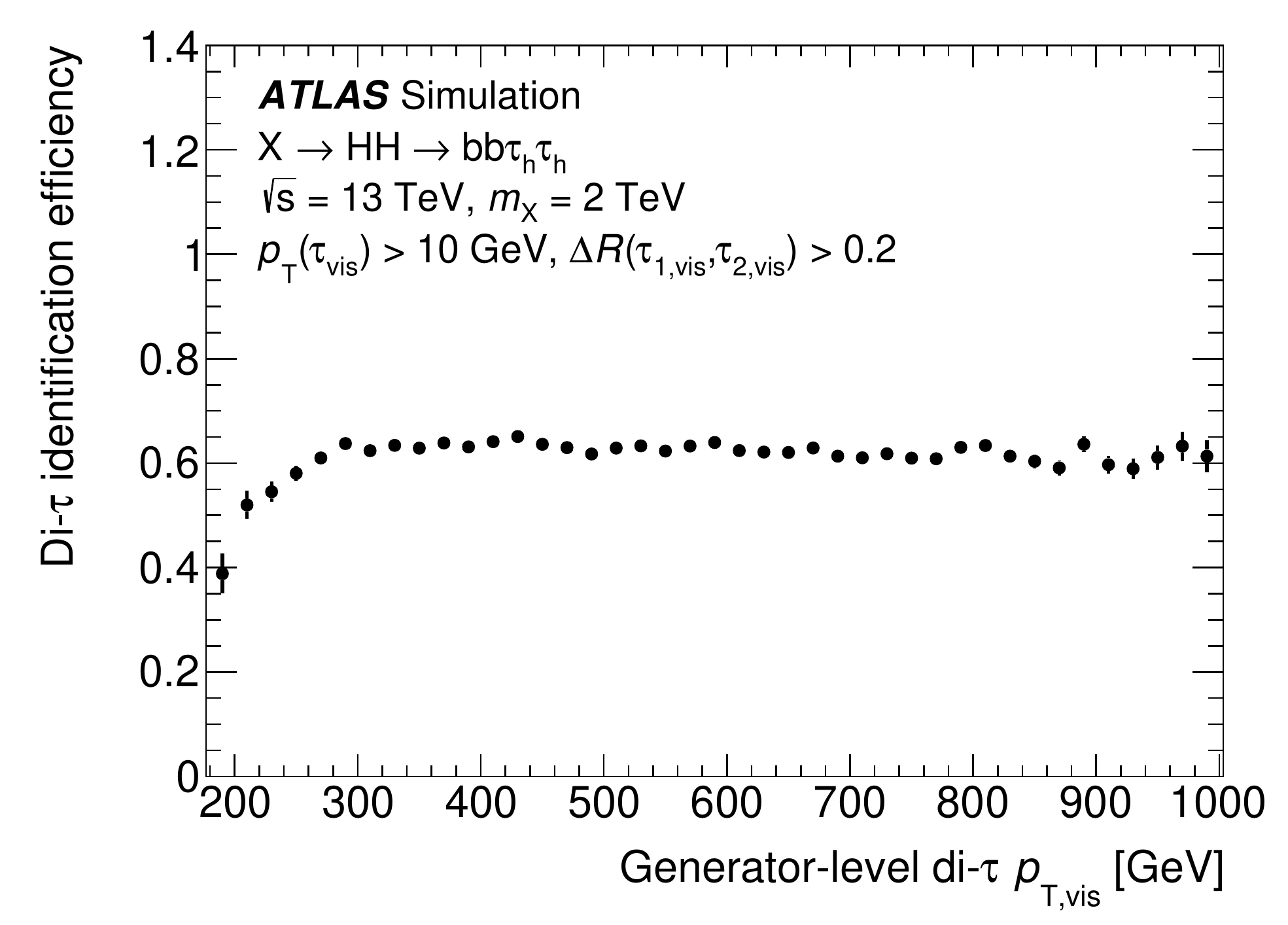}\label{fig:ideffvspt}
}
\subfloat[]{
\includegraphics[width=0.45\textwidth]{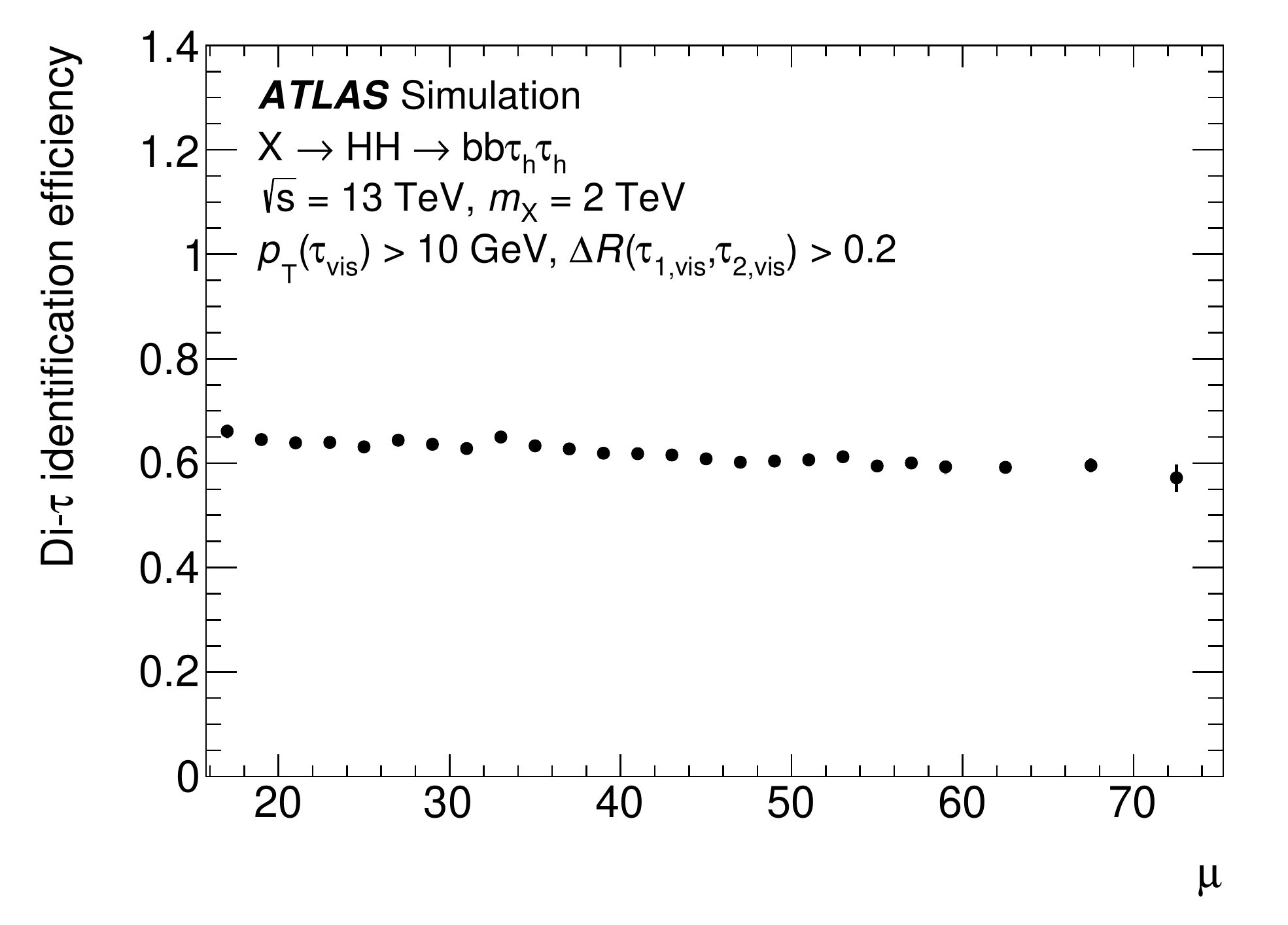}\label{fig:ideffvsmu}
}
\caption{
(a) Background rejection factor versus identification efficiency for the BDT used to discriminate between
di-$\tau$ signatures and quark- or gluon-initiated jets, as obtained after the di-$\tau$ reconstruction step
(the red cross indicates the benchmark used in the analysis, corresponding to a signal efficiency of about 60\%
and a background rejection factor of $10^4$), and identification efficiency of boosted di-$\tau$ objects,
as a function of (b) their \pT\ at generator level and (c) the number $\mu$ of pile-up interactions. The
vertical error bars only account for statistical uncertainties.
}
\label{fig:bdt}
\end{figure}
 
Since the majority of hadronically decaying $\tau$-leptons produce one or three charged pions, the number of
tracks matched to each di-$\tau$ sub-jet, with a distance $\Delta R < 0.1$ from its axis, must be either
one or three in the following. The efficiency of this additional requirement, computed after the di-$\tau$
reconstruction and identification steps, is shown in Figures~\ref{fig:trkeffvspt} and~\ref{fig:trkeffvsmu}
as a function of the generator-level \pT of the di-$\tau$ system and the number $\mu$ of pile-up interactions,
respectively. For this purpose, the same sample of simulated $X \to HH \to b\bar{b}\tau^+\tau^-$ events
is used. The di-$\tau$ sub-jet track selection is found to be stable not only with respect
to the \pT of the di-$\tau$ system and the number of pile-up interactions, but also with respect to pseudorapidity
and the distance $\Delta R(\tau_{1,\text{vis}},\tau_{2,\text{vis}})$ between the visible decay products of the
two hadronically decaying $\tau$-leptons. This additional requirement further improves the rejection of the
background with quark- or gluon-initiated jets by a factor of about five.
 
\begin{figure}[htbp]
\centering
\subfloat[]{
\includegraphics[width=0.48\textwidth]{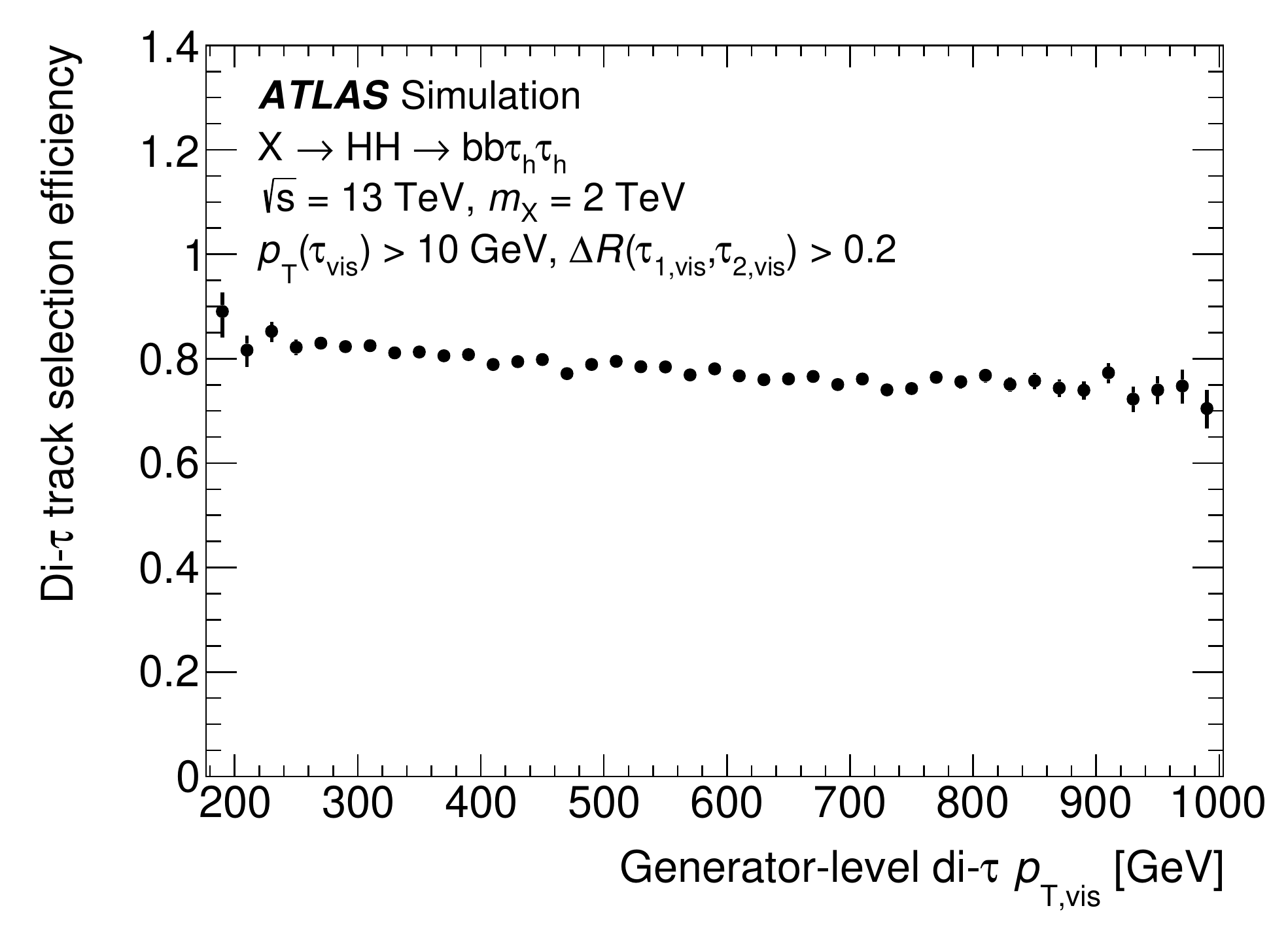}\label{fig:trkeffvspt}
}
\subfloat[]{
\includegraphics[width=0.48\textwidth]{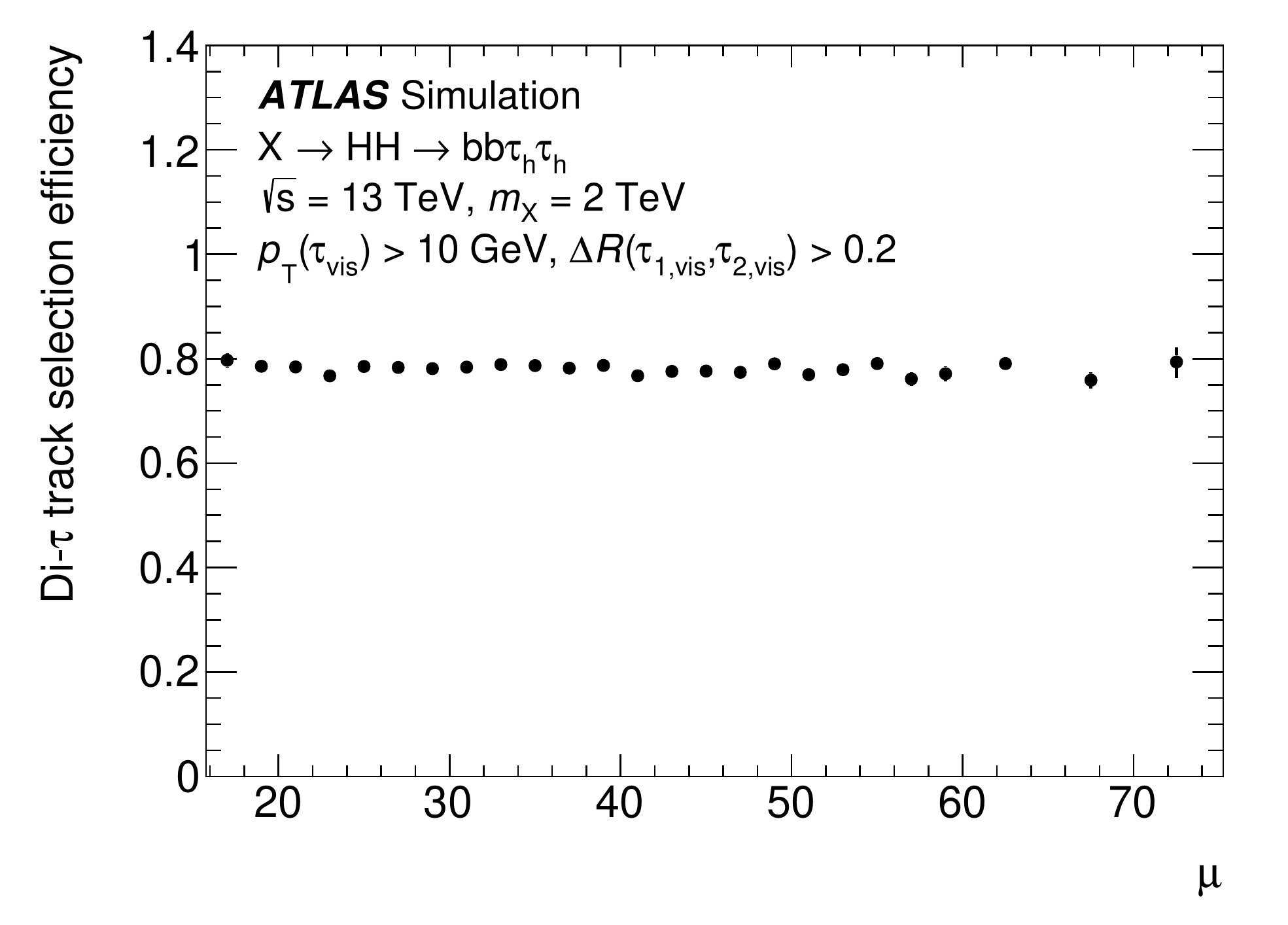}\label{fig:trkeffvsmu}
}
\caption{Efficiency of the track selection for di-$\tau$ sub-jets versus (a) the \pT\ of the di-$\tau$ system
at generator level and (b) the number $\mu$ of pile-up interactions, computed after the boosted
di-$\tau$ reconstruction and identification steps in simulated $X \to HH \to b\bar{b}\tau^+\tau^-$
events, where the resonance mass is set to 2\,\TeV\ and both $\tau$-leptons decay hadronically.
The vertical error bars only account for statistical uncertainties.
}
\label{fig:track-eff}
\end{figure}
 
In simulated $X \to HH \to b\bar{b}\tau^+\tau^-$ events, the energy of reconstructed di-$\tau$ objects is close
to that of the visible decay products of the corresponding two hadronically decaying $\tau$-leptons at generator
level, as illustrated in Figure~\ref{fig:ditau-ene-pt}. In addition, the energy resolution is found to improve with
increasing energy. In regions of the data enriched in boosted di-$\tau$ candidates, good agreement is found
between simulated and measured di-$\tau$ energy distributions; hence no additional energy calibration is applied
to di-$\tau$ objects.
 
\begin{figure}[htbp]
\centering
\subfloat[]{
\includegraphics[width=0.48\textwidth]{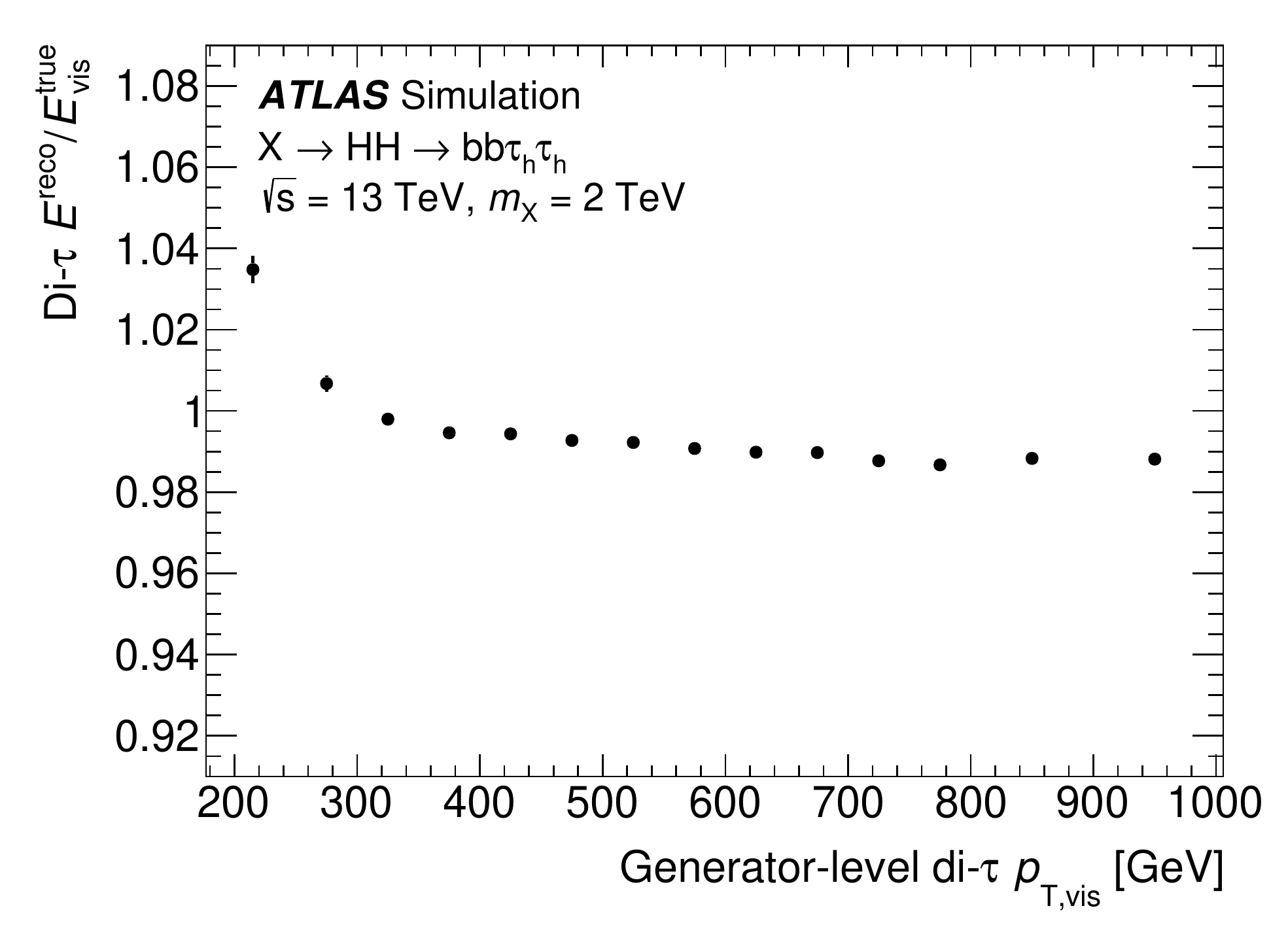}\label{fig:e1-pt}
}
\subfloat[]{
\includegraphics[width=0.48\textwidth]{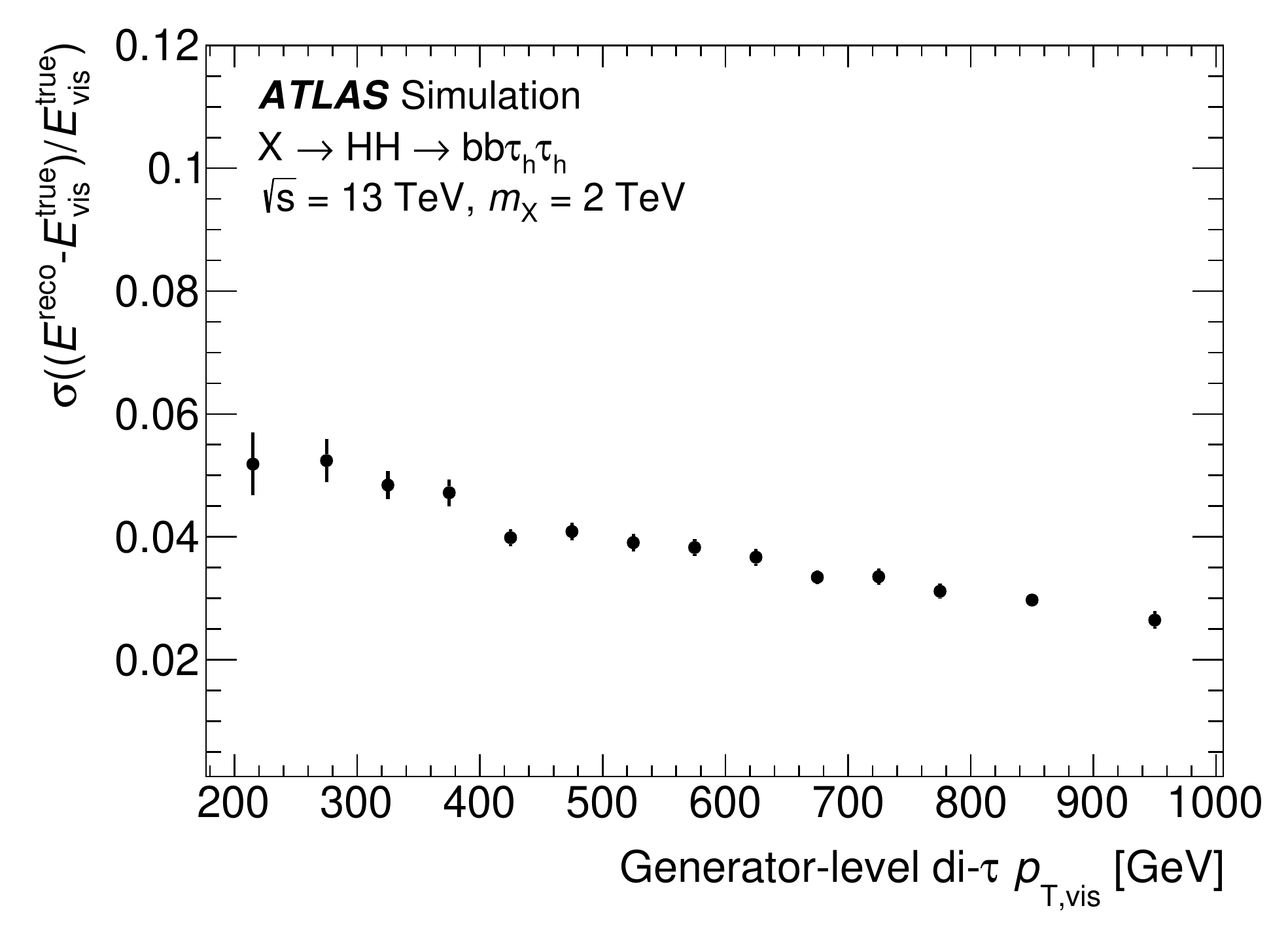}\label{fig:e2-pt}
}
\caption{
Energy (a) scale and (b) resolution of the di-$\tau$ reconstruction as a function of the \pT\ of the di-$\tau$
system at generator level, computed in simulated $X \to HH \to b\bar{b}\tau^+\tau^-$ events, where the
resonance mass is set to 2\,\TeV\ and all $\tau$-leptons decay hadronically. The requirement of $\pT > 300\,\GeV$
on the seeding jet truncates the reconstructed energy distribution and leads to a degradation of the energy scale
for the lower values of the \pT of the di-$\tau$ system. The vertical error bars only account for statistical
uncertainties.
}
\label{fig:ditau-ene-pt}
\end{figure}
 
 
\section{Event selection and categorisation}
\label{sec:evtselection}
This section presents the common event selection and further categorisations used for the multi-jet
background estimation, the data-driven correction of the efficiency of the di-$\tau$ tagger and the search
for $X \to HH \to b\bar{b}\tau^+\tau^-$. The events are selected using unprescaled triggers requiring the presence
of a large-radius jet with an \ET\ threshold between 360 and 460\,\GeV, depending on the data-taking period.
Subsequently, in order to ensure a constant trigger efficiency, an offline \pT\ cut that exceeds the trigger
threshold by 40--50\,\GeV\ is applied to the leading large-radius jet, independently of whether it is
compatible with a boosted $b\bar{b}$ or hadronically decaying $\tau^+\tau^-$ pair. In order to avoid
contamination from non-collision backgrounds, such as those originating from calorimeter noise, beam halo and
cosmic rays, events are rejected if they contain an \antikt\ jet with $R = 0.4$ and $\pT > 20\,\GeV$ that does
not fulfil the loose quality criteria of Ref.~\cite{ATLAS-CONF-2015-029}. In addition, events are required to
have a vertex with at least two associated tracks with $\pT > 500\,\MeV$. Finally, events are vetoed if they
contain an electron or muon that meets the requirements described in Section~\ref{sec:objects}.
 
At preselection, events are required to contain at least one reconstructed
di-$\tau$ object, which must meet the requirements below, in addition to those listed in Section~\ref{sec:ditau}.
If multiple di-$\tau$ objects are found, the one with the highest \pT\ is selected.
\begin{itemize}
\item The number of sub-jets is at most three, and the \pT\ threshold of the two leading sub-jets
is raised to 50\,\GeV\ in order to reduce the contribution of multi-jet background events.
\item The separation $\Delta R$ between the leading and sub-leading sub-jets is less than 0.8 to ensure
that both are fully contained in the di-$\tau$ seeding jet.
\item The charge product of the two leading sub-jets is $Q = q^\text{lead}\times q^\text{sub-lead} = \pm 1$.
The charge of the sub-jet is defined as the sum of the charges of the associated tracks.
\item The transverse momentum of the di-$\tau$ object, computed from the two leading sub-jets, must
exceed 300\,\GeV\ (or fulfil the tighter \pT\ requirement set by the trigger threshold if the seeding large-radius
jet fires the trigger), with a pseudorapidity $|\eta| < 1.37$ or $1.52 < |\eta| < 2.0$, thereby rejecting the
transition region between the barrel and endcap calorimeters.
\end{itemize}
 
In addition, $\met > 10\,\GeV$ is required in order to define a direction for the missing transverse momentum.
To select boosted $H \to b\bar{b}$ decays, large-$R$ jets with a \pT\ exceeding 300\,\GeV\ (or the tighter
trigger-dependent threshold if the seeding large-radius jet fires the trigger), $|\eta|<2.0$,
a combined mass larger than 50\,\GeV\
and a separation $\Delta R > 1.0$ from the selected di-$\tau$ object are required. If multiple
large-$R$ jets fulfil these requirements, the one with the highest \pT\ is selected.
 
Events are then further divided into regions with either misidentified or true di-$\tau$ objects.
In the latter case, the event categories contain a significant fraction of di-$\tau$ objects from
either $Z \to \tau^+\tau^-$ or $H \to \tau^+\tau^-$ decays.
As shown in Table~\ref{tab:regions}, the categorisation is based on the charge product $Q$ of the two
leading di-$\tau$ sub-jets, the number $N_{b\text{-tags}}$ of $b$-tagged track-jets in the selected
large-$R$ jet, the azimuthal angle $|\Delta \phi|$ between the selected di-$\tau$
object and the missing transverse momentum, the combined mass $m_J$ of the selected large-$R$ jet,
and the invariant mass of the selected di-$\tau$ object and the large-$R$ jet, $m_{HH}^\text{vis}$.
Requirements on $m_J$ and $m_{HH}^\text{vis}$ are imposed in regions with one $b$-tagged track-jet
in order to suppress the contamination from $X \to HH \to b\bar{b}\tau^+\tau^-$ signal events. Finally,
in order to ensure orthogonality to searches for $X \to HH \to b\bar{b}b\bar{b}$~\cite{EXOT-2016-31},
events with two or more $b$-tagged large-$R$ jets are removed.
 
\begin{table}[htbp]
\centering
\caption{Definition of various categories after the event preselection (see text for details), based on
the charge product $Q$ of the two leading di-$\tau$ sub-jets (OS for opposite-sign, SS for same-sign),
the number $N_{b\text{-tags}}$ of $b$-tagged track-jets in the selected large-$R$ jet, the
azimuthal angle $|\Delta\phi|$ between the di-$\tau$ object and the missing transverse momentum,
the large-$R$ jet mass $m_J$, and the visible mass $m_{HH}^\text{vis}$ of the reconstructed $HH$ system.}
\label{tab:regions}
\small
\begin{tabular}{llccccc}
\toprule
Name & Usage  & $Q$ & $N_{b\text{-tags}}$ & $|\Delta\phi|$ & $m_J$\,[\GeV] & $m_{HH}^\text{vis}$\,[\GeV] \\
\midrule
FF SS             &  Fake-factor (FF) computation    & $+1$ & 0      & -    & -                      & - \\
MJ OS 0-tag       &  Closure test, multi-jet 0-$b$-tag & $-1$ & 0      & $>1$ & -                      & - \\
MJ OS 1-tag       &  Closure test, multi-jet 1-$b$-tag & $-1$ & 1      & $>1$ & 50--60 or $>160$  & $<1500$ \\
\midrule
Z$\tau\tau$ 0-tag &  Di-$\tau$ tagger correction  & $-1$ & 0      & $<1$ & -                      & - \\
Z$\tau\tau$ 1-tag &  Di-$\tau$ tagger closure test   & $-1$ & 1      & $<1$ & 50--60 or $>160$  & $<1500$ \\
\midrule
$HH$ signal region  &  $X \to HH \to b\bar{b}\tau^+\tau^-$ search & $-1$ & 2 & $<1$ & $>60$ and $<160$ & $>[0, 900, 1200]$ \\
\bottomrule
\end{tabular}
\end{table}
 
\section{Estimation of the multi-jet background with a misidentified di-$\tau$ object}
\label{sec:ddbkgest}
Multi-jet events with quark- or gluon-initiated jets misidentified as di-$\tau$ objects are a
common background for the efficiency measurement of the di-$\tau$ tagger and in the search for
$X \to HH \to b\bar{b}\tau^+\tau^-$. This section describes the data-driven method developed
to estimate this background, whereas non-multi-jet background events with a misidentified
di-$\tau$ object are obtained directly from simulation.
 
A data sample enriched in misidentified di-$\tau$ objects is collected using the same event selection
as in a region of interest (RoI), except that the selected di-$\tau$ object must fail (instead of pass)
the BDT-based identification, but it still fulfils a very loose criterion that corresponds to a cut on
the BDT output score with an efficiency above 99\%. In fact, this very loose criterion is applied to all
di-$\tau$ candidates in order to keep the composition of misidentified di-$\tau$ objects in terms of
quarks and gluons close to that in the RoI. The contribution from non-multi-jet processes, obtained
using simulation, is subtracted from the data. The remaining events, defined as $N^\text{fail}_\text{RoI}$,
are then reweighted by a fake factor (FF) to predict the yield of events $N^\text{mis-ID}_\text{RoI}$ with
a misidentified di-$\tau$ object in the RoI as follows:
\begin{linenomath*}
\begin{equation*}
N^\text{mis-ID}_\text{RoI} = N^\text{fail}_\text{RoI} \times \text{FF}.
\label{eq:nfakes}
\end{equation*}
\end{linenomath*}
 
The FF is calculated in the region labelled as `FF SS' in Table~\ref{tab:regions}, which requires $Q = +1$ for the
di-$\tau$ object and is enriched in multi-jet events. It is binned in the \pt\ of the di-$\tau$ object and defined as:
\begin{linenomath*}
\begin{equation*}
\text{FF} = \frac{ N^\text{pass}_\text{FF SS} }{ N^\text{fail}_\text{FF SS} },
\label{eq:ff}
\end{equation*}
\end{linenomath*}
where the numerator is the number of events passing the BDT-based di-$\tau$ identification requirement, while
the denominator is the number of events failing this requirement but passing the very loose BDT requirement.
The contributions from non-multi-jet processes in the numerator and the denominator are about 24\% and 3\% of
the total event yields, respectively, and they are subtracted. Figure~\ref{fig:ff} shows the FF as a function of
the \pT\ of the di-$\tau$ object. The observed increase of the FF relates to a reduced rejection of the background
from quark- or gluon-initiated jets at higher \pT\ values of the di-$\tau$ object. On the other hand, the FF is
inclusive in $\eta$, since no significant change in the multi-jet background modelling was found with an additional
binning in the pseudorapidity of the di-$\tau$ object.
The statistical uncertainty of the measured FF comes from the limited size of the samples of data and simulated
events used in the computation.
 
\begin{figure}[htbp]
\begin{center}
\includegraphics[width=0.55\textwidth]{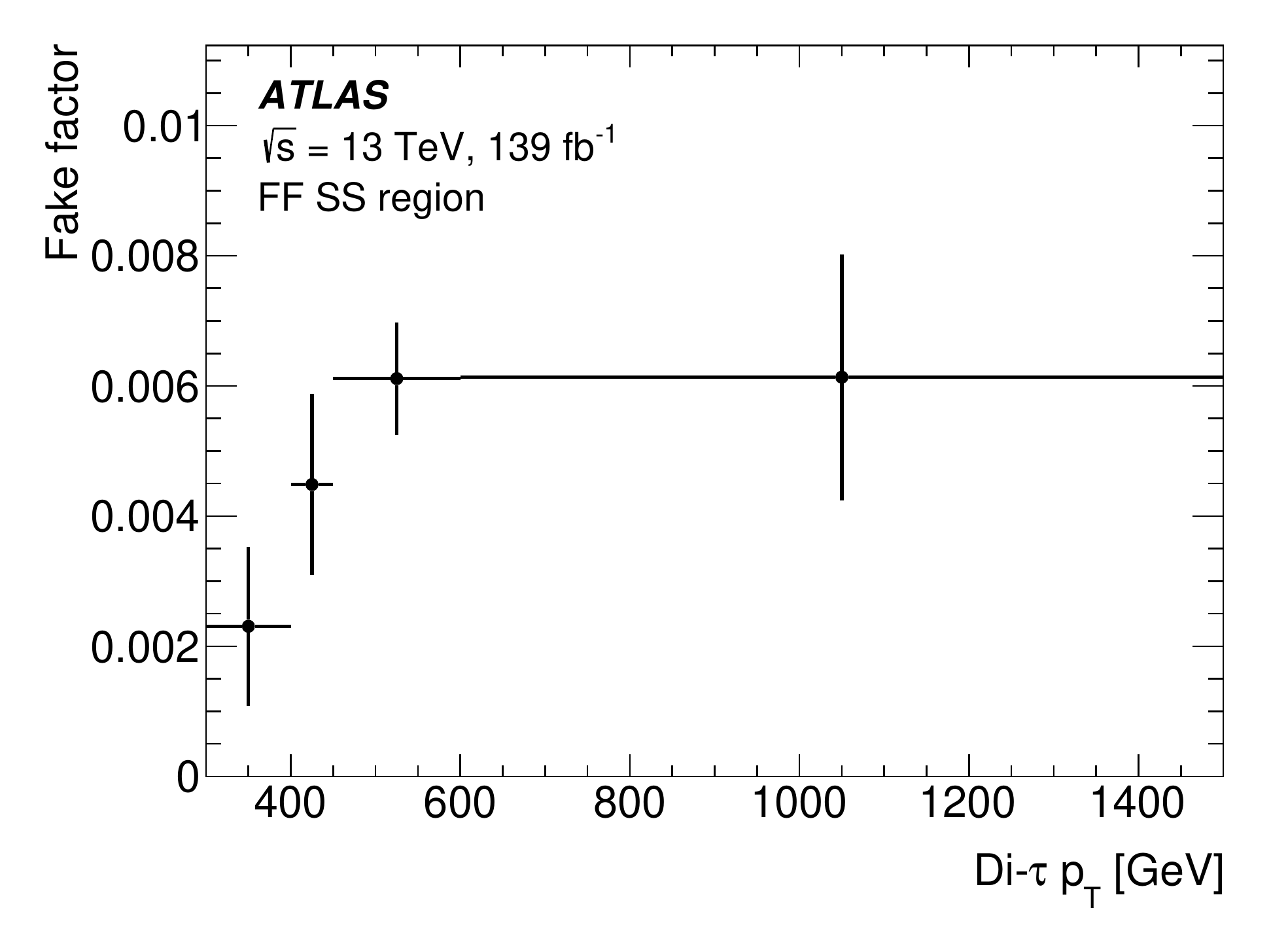}
\end{center}
\caption{Fake factor as a function of the \pT\ of the di-$\tau$
object, computed in the control region labelled as `FF SS' in Table~\ref{tab:regions},
with the requirements of $Q= +1$ for the di-$\tau$ object and of zero $b$-tagged track-jets
for the selected large-$R$ jet. The error bars indicate the statistical uncertainties.}
\label{fig:ff}
\end{figure}

Systematic uncertainties are computed to account for the different event selections used in the analysis.
In particular, a \pt-binned FF is computed in a region with the same selections as those used in `FF SS',
except that the charge requirement for the di-$\tau$ candidate is changed to $Q = -1$. The deviation
(up to 30\%) from the nominal FF is taken as a systematic uncertainty in the extrapolation from the
SS region to the OS region.
In addition, the maximum overall difference (42\%) between the FF values evaluated in regions with zero,
one or two $b$-tagged track-jets in the selected large-$R$ jet accounts for the systematic uncertainty
in the event categorisation based on $b$-tagging.
 
The estimation of the multi-jet background is validated in two regions (`MJ OS 0-tag' and `MJ OS 1-tag'
in Table~\ref{tab:regions}) selected by requiring $Q= -1$ for the di-$\tau$ candidate, either zero or one
$b$-tagged track-jet in the selected large-$R$ jet, and $|\Delta \phi| > 1$ between the selected
di-$\tau$ object and the missing transverse momentum to suppress events with correctly identified di-$\tau$
objects. As shown in Figure~\ref{fig:qcd-os-vr}, good agreement between the predicted and measured multi-jet
backgrounds is found in the `MJ OS 0-tag' region, while some discrepancy is observed between the predicted
and measured event yields in the `MJ OS 1-tag' region.\footnote{The simulated background events with a true
di-$\tau$ object are corrected to account for the measured efficiency of the di-$\tau$ tagger, as discussed in
Section~\ref{sec:ditau-calib}, and an additional normalisation factor is subsequently applied to correct
for the modelling of the $Z$+hf background, as discussed in Section~\ref{sec:hhbbtautau}.}
Therefore, an additional, conservative, non-closure systematic uncertainty of 50\% is assigned for the
multi-jet background estimation in all regions with $N_{b\text{-tags}}\ge 1$.
 
\begin{figure}[htbp]
\centering
\subfloat[]{
\includegraphics[width=0.45\textwidth]{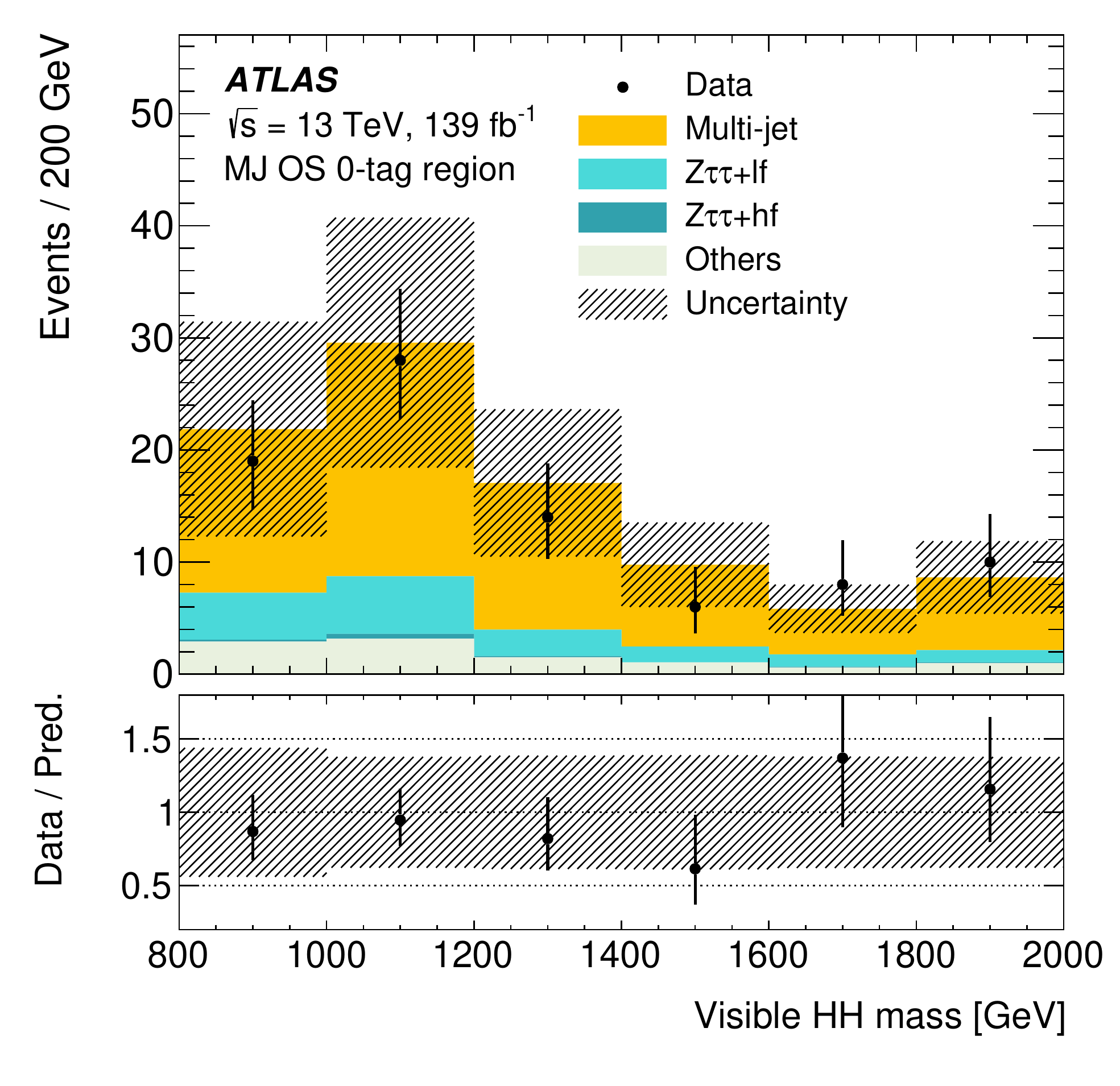}
}
\subfloat[]{
\includegraphics[width=0.45\textwidth]{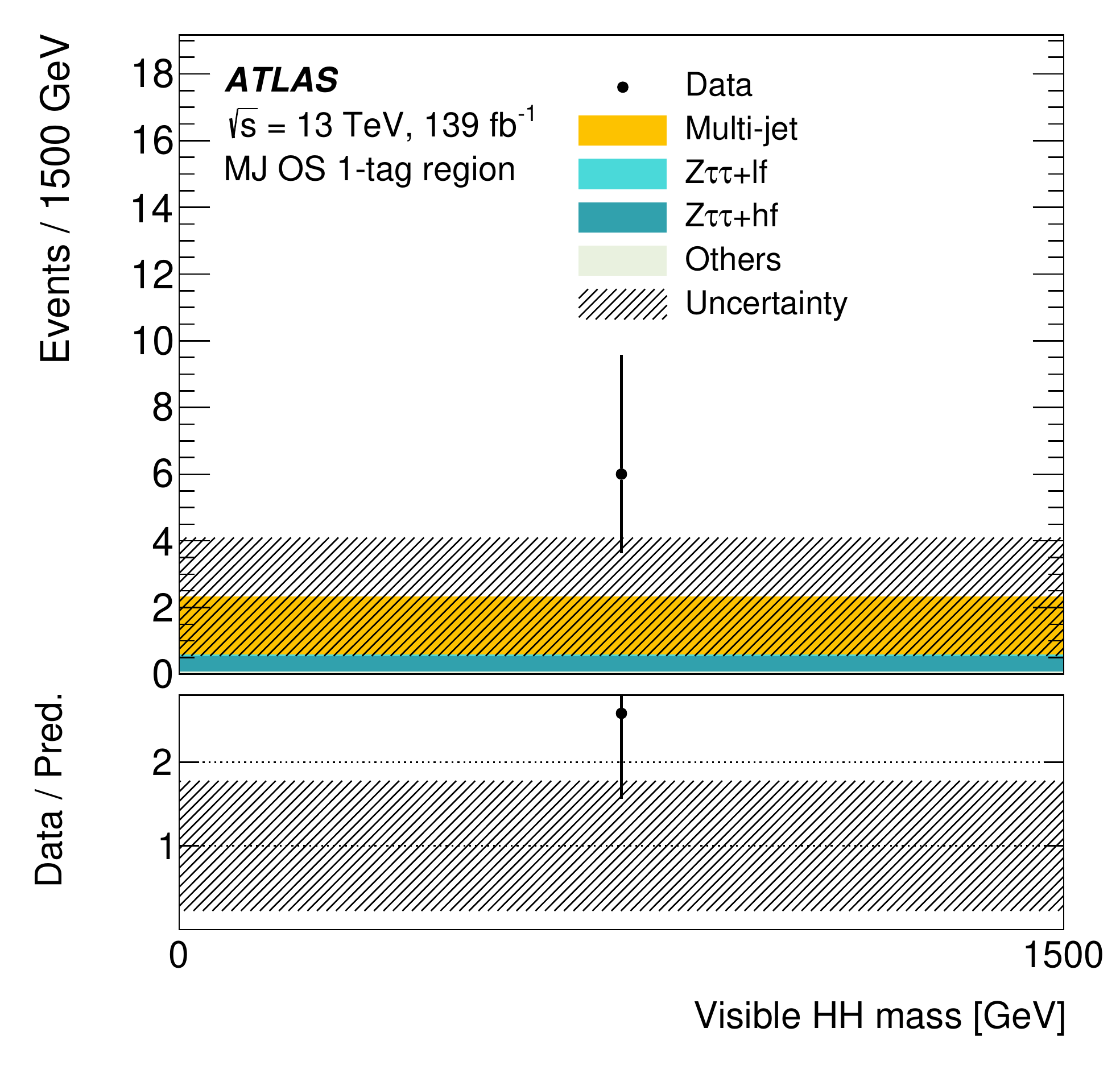}
}
\caption{Distribution of $m_{HH}^\text{vis}$ in two regions enriched in multi-jet events,
(a) `MJ OS 0-tag' and (b) `MJ OS 1-tag' in Table~\ref{tab:regions}. The background
referred to as `Others' contains $W$+jets, diboson, $ZH$, \ttbar\ and single-top-quark events.
The lower panels show a bin-by-bin comparison of the data and the predicted total background,
in terms of event yield ratios.
The hatched bands represent combined statistical and systematic uncertainties.}
\label{fig:qcd-os-vr}
\end{figure}
 
 
\section{Data-driven correction of the di-$\tau$ tagger efficiency}
\label{sec:ditau-calib}
To account for differences in the efficiency of the di-$\tau$ tagger between simulation
and data, a correction is derived in a region enriched with properly identified
di-$\tau$ objects from boosted $Z \to \tau^+\tau^-$ decays.
This region, labelled as `Z$\tau\tau$ 0-tag' in Table~\ref{tab:regions}, is designed by imposing
a veto on events with $b$-tagged track-jets in the selected large-$R$ jet and by demanding that $Q=-1$
and that the boosted di-$\tau$ system and the missing transverse momentum point in the same direction,
i.e.\ $|\Delta \phi| < 1$. Figure~\ref{fig:Z-calib-CR-0b} shows the visible di-$\tau$ mass in this region,
as well as \pT and $\eta$ distributions after an additional requirement on the visible di-$\tau$ mass,
$30\,\GeV < m_{\text{di-}\tau}^{\text{vis}} < 90\,\GeV$, which increases the fraction of $Z \to \tau^+\tau^-$
events to 87\%.\footnote{The contribution of $Z \to \tau^+\tau^-$ events in which one of the $\tau$-leptons
decays to an electron or muon is negligible. A muon rarely deposits enough energy in the calorimeter to be
reconstructed as a sub-jet of the di-$\tau$ object. On the other hand, the di-$\tau$ reconstruction and
identification algorithms provide only little discrimination against $Z \to \tau^+\tau^-$ events with an
electron in the decay chain. However, the lepton veto, which excludes non-isolated electrons, ensures that
such events only have a minor contribution.}
 
\begin{figure}[htbp]
\centering
\subfloat[]{
\includegraphics[width=0.45\textwidth]{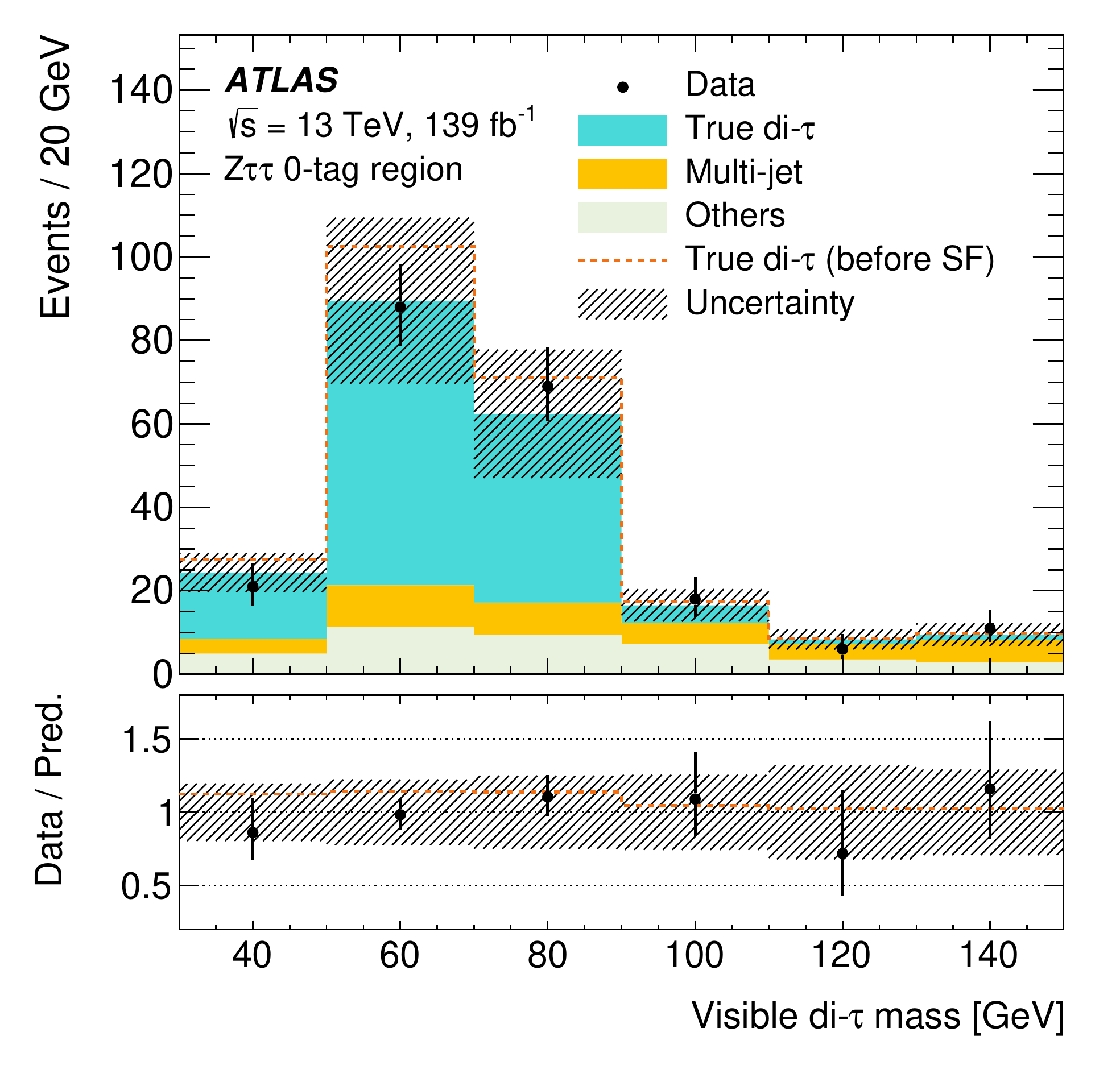} \label{fig:Z-calib-CR-0b-a}
}\\
\subfloat[]{
\includegraphics[width=0.45\textwidth]{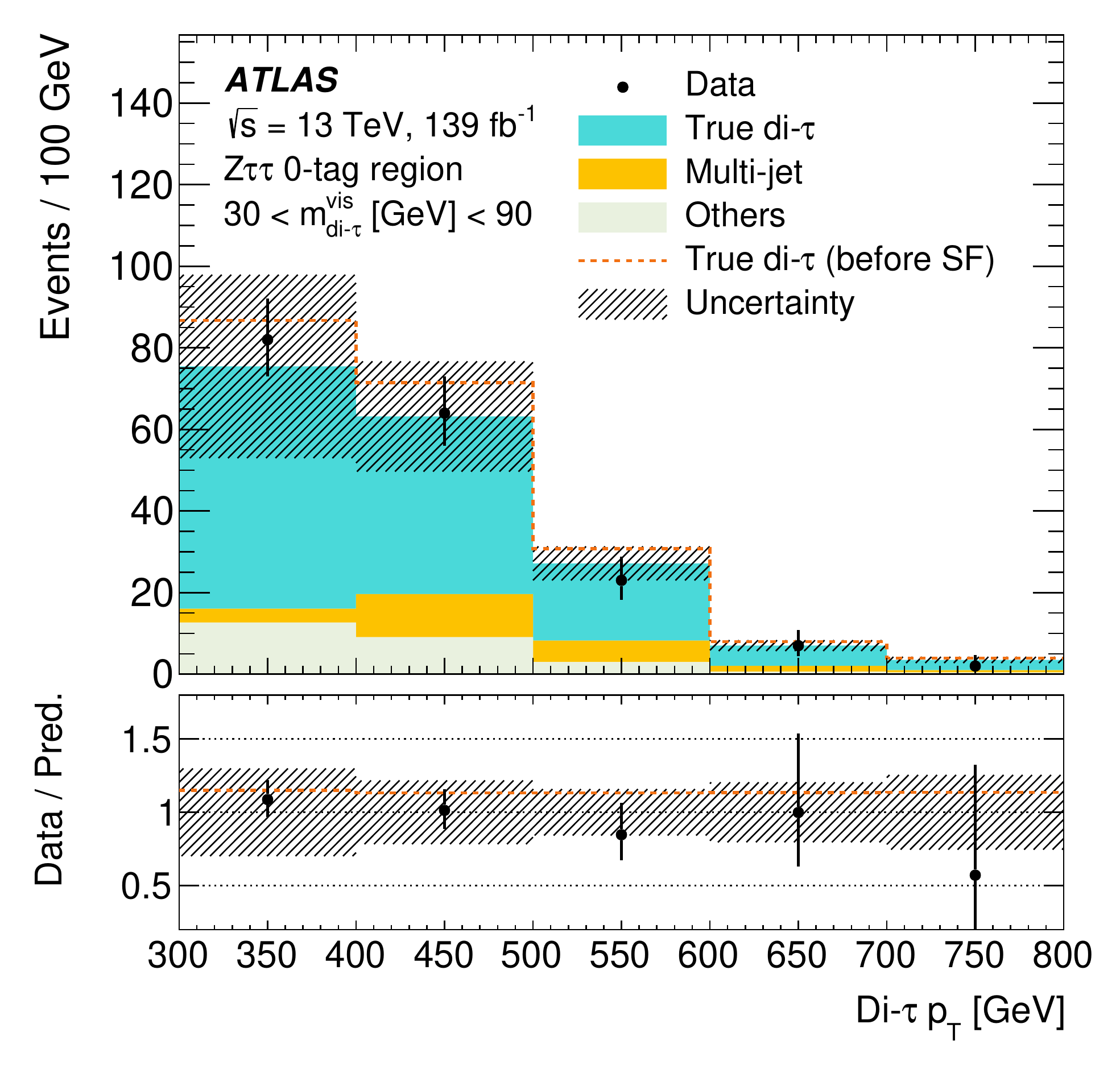} \label{fig:Z-calib-CR-0b-b}
}
\subfloat[]{
\includegraphics[width=0.45\textwidth]{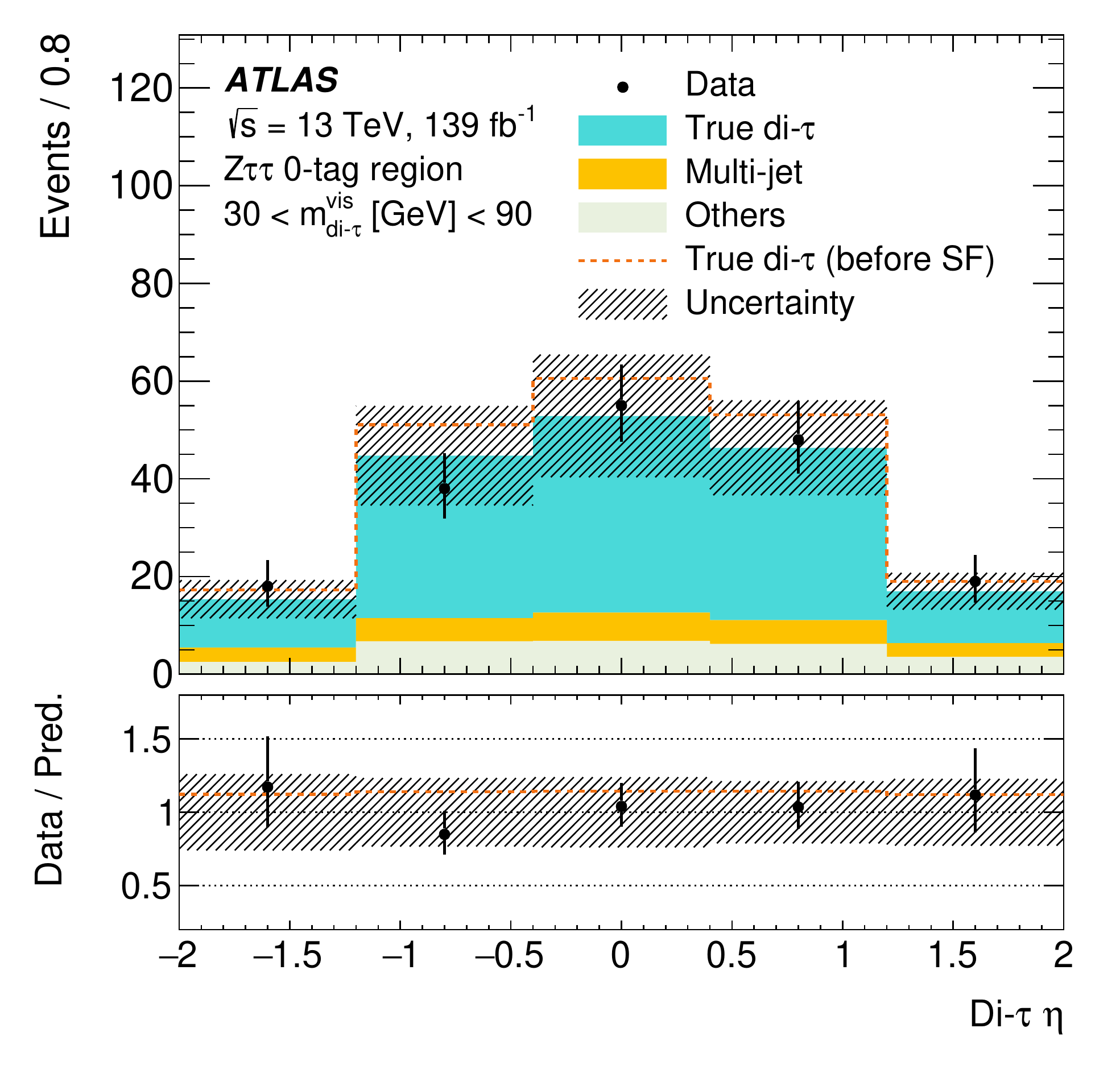} \label{fig:Z-calib-CR-0b-c}
}
\caption{Distributions of (a) the visible mass of the di-$\tau$ object, (b) its \pT and (c) $\eta$ in
the region labelled as `Z$\tau\tau$ 0-tag' in Table~\ref{tab:regions}. All simulated events containing
a generator-level $\tau^+\tau^-$ pair matched to a simulated di-$\tau$ object are referred to as
`True di-$\tau$'.
The lower panels show a bin-by-bin comparison of the data and the predicted total background,
in terms of event yield ratios.
The hatched bands represent combined statistical and systematic uncertainties.}
\label{fig:Z-calib-CR-0b}
\end{figure}
 
A scale factor (SF) is computed in the `Z$\tau\tau$ 0-tag' region as the ratio of measured to predicted
event yields with a true boosted hadronically decaying $\tau^+\tau^-$ pair. The measured event yield is
obtained after subtracting the backgrounds with a misidentified di-$\tau$ object from the data:
\begin{linenomath*}
\begin{equation*}
\mbox{SF} = \frac{N(\text{data}) - N(\text{non-di-}\tau)}{N(\text{true di-}\tau)} = 0.84 \pm 0.09 \text{ (stat) }
^{+0.14}_{-0.13}\text{ ($Z$-modelling) }
^{+0.19}_{-0.20}\text{ (syst)}.
\end{equation*}
\end{linenomath*}
No significant dependence of the ratio on the \pT\ or $\eta$ of the di-$\tau$ object is found within the
uncertainties, as illustrated in Figures~\ref{fig:Z-calib-CR-0b-b} and~\ref{fig:Z-calib-CR-0b-c}.  Hence, a
single-bin SF is considered in order to minimise the statistical uncertainty. The SF is applied to all simulated
(background and signal) events containing a di-$\tau$ object matched to a hadronically decaying $\tau^+\tau^-$
pair at generator level.
 
The following uncertainties are considered in the SF computation. The statistical uncertainty accounts for
the limited size of the data and simulated background samples, and is respectively 10.3\% and 2.7\% relative
to the SF. The uncertainty labelled as `$Z$-modelling' arises from the cross-section (5\%) and the acceptance
of $Z$+jets events. The latter comes from the choice of PDF set, as well as from variations of $\alphas$ and
of the renormalisation, factorisation, matrix-element matching and soft-gluon resummation scales in \textsc{Sherpa}.
The uncertainty related to the cross-section and acceptance of the other
simulated processes is found to be negligible. The systematic uncertainties arising from the multi-jet
background estimate are discussed in Section~\ref{sec:ddbkgest}. In all samples of simulated events, the
instrumental systematic uncertainties arise primarily from the mismodelling of large-$R$ jets, in particular
the jet energy scale and resolution, as well as the jet mass scale and resolution~\cite{JETM-2018-02}.
Uncertainties arising from the mismodelling of the reconstructed energy (momentum) measurements for
electrons (muons), as well as from the mismodelling of their reconstruction and identification
efficiencies~\cite{EGAM-2018-01,PERF-2015-10}, are found to be negligible. The systematic uncertainties
arising from the \met scale and resolution~\cite{PERF-2016-07}, as well as from the mismodelling
of pile-up, are taken into account. Finally, the uncertainty in the combined 2015--2018 integrated
luminosity is 1.7\%~\cite{ATLAS-CONF-2019-021}, obtained using the LUCID-2 detector~\cite{LUCID2}
for the primary luminosity measurements.
 
The modelling of di-$\tau$ objects from genuine hadronically decaying $\tau^+\tau^-$ pairs is checked
in the region `Z$\tau\tau$ 1-tag' with exactly one $b$-tagged track-jet in the selected large-$R$
jet. In the following, for events containing a large-$R$ jet with $b$-tagged track-jets, an additional
systematic uncertainty arises from differences between the predicted and measured $b$-tagging efficiency
and rates at which $c$-jets and light-flavour jets are misidentified as $b$-tagged
jets~\cite{FTAG-2018-01, ATLAS-CONF-2018-001, ATLAS-CONF-2018-006}. As shown in
Figure~\ref{fig:Z-calib-VR-1b}, good agreement between predicted and measured
yields is observed in the region `Z$\tau\tau$ 1-tag', within the uncertainties.
 
\begin{figure}[htbp]
\centering
\includegraphics[width=0.45\textwidth]{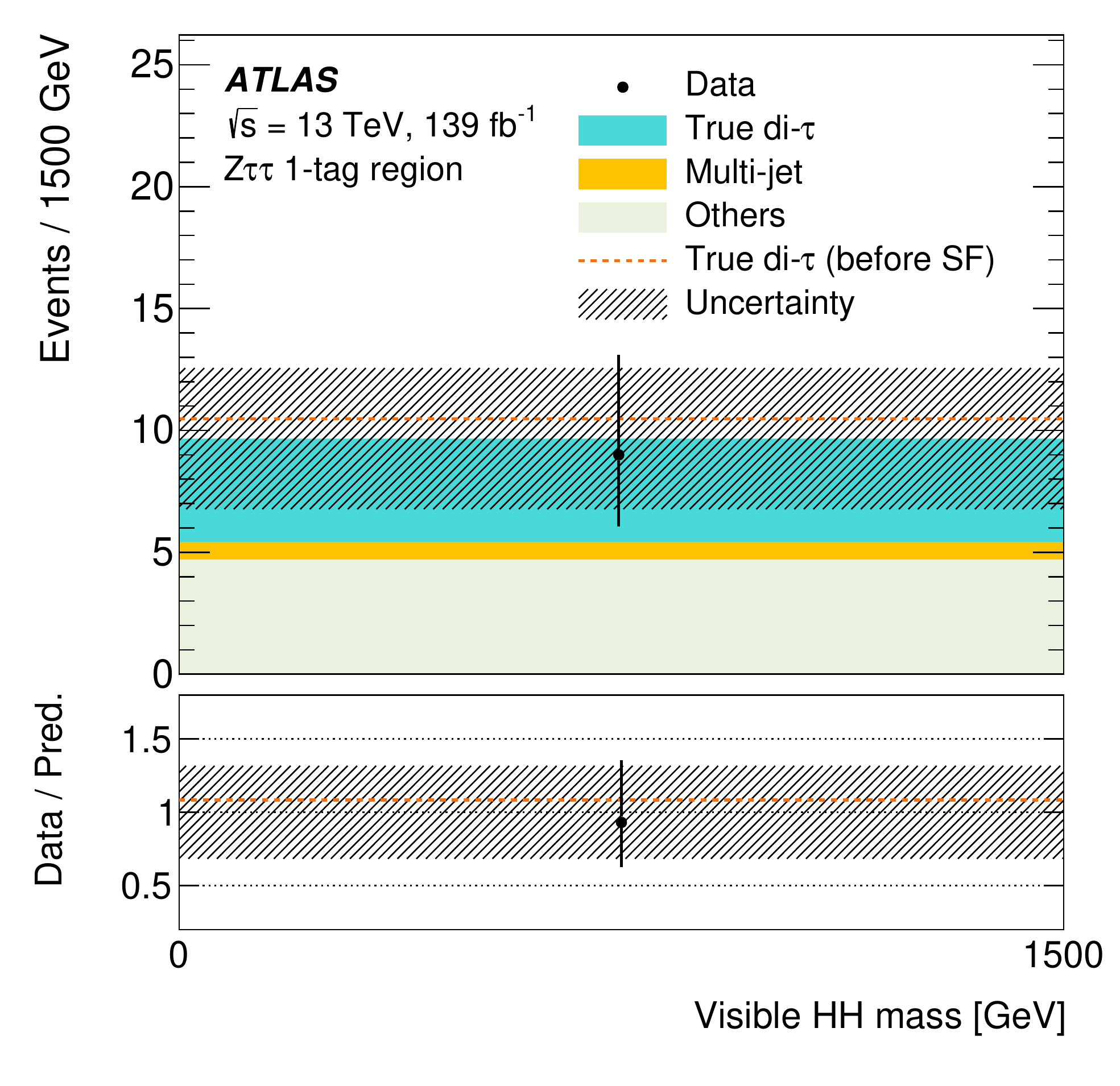}
\caption{Predicted and measured event yields in the region labelled as `Z$\tau\tau$ 1-tag'
in Table~\ref{tab:regions}. All simulated events containing a generator-level $\tau^+\tau^-$ pair
matched to a simulated di-$\tau$ object are referred to as `True di-$\tau$'.
The main contribution to the background referred to as ‘Others’ comes from the \ttbar\ process (about 3.6 events).
The lower panel shows the event yield ratio of the data and the predicted total background.
The hatched bands represent combined statistical and systematic uncertainties.}
\label{fig:Z-calib-VR-1b}
\end{figure}
 
\section{Search for resonant Higgs boson pair production in the $b\bar{b}\tau^+\tau^-$ final state}
\label{sec:hhbbtautau}
The signal region (SR) used in the search for resonant Higgs boson pair production is defined by requiring
OS charges for the two leading sub-jets of the di-$\tau$ object ($Q=-1$), $|\Delta \phi| < 1$, and
two $b$-tagged track-jets in the selected large-$R$ jet, which furthermore must have a mass $m_J$
compatible with the mass of the Higgs boson (60--160\,\GeV). Finally, a requirement on $m_{HH}^\text{vis}$
is imposed, which depends on the resonance mass hypothesis: no requirement is applied if $m_X < 1.6\,\TeV$,
whereas the cut-off value is set to 900 (1200)\,\GeV\ if $m_X \geq 1.6\,(2.5)\,\TeV$. These requirements are
summarised in the last row of Table~\ref{tab:regions}. The signal acceptance times efficiency at various stages
of the event selection is shown in Figure~\ref{fig:fig_eff_acc_1}. The efficiency at low resonance masses is
limited by the trigger thresholds and kinematic requirements on the large-$R$ jet.
 
\begin{figure}[htbp]
\begin{center}
\includegraphics[width=0.5\textwidth]{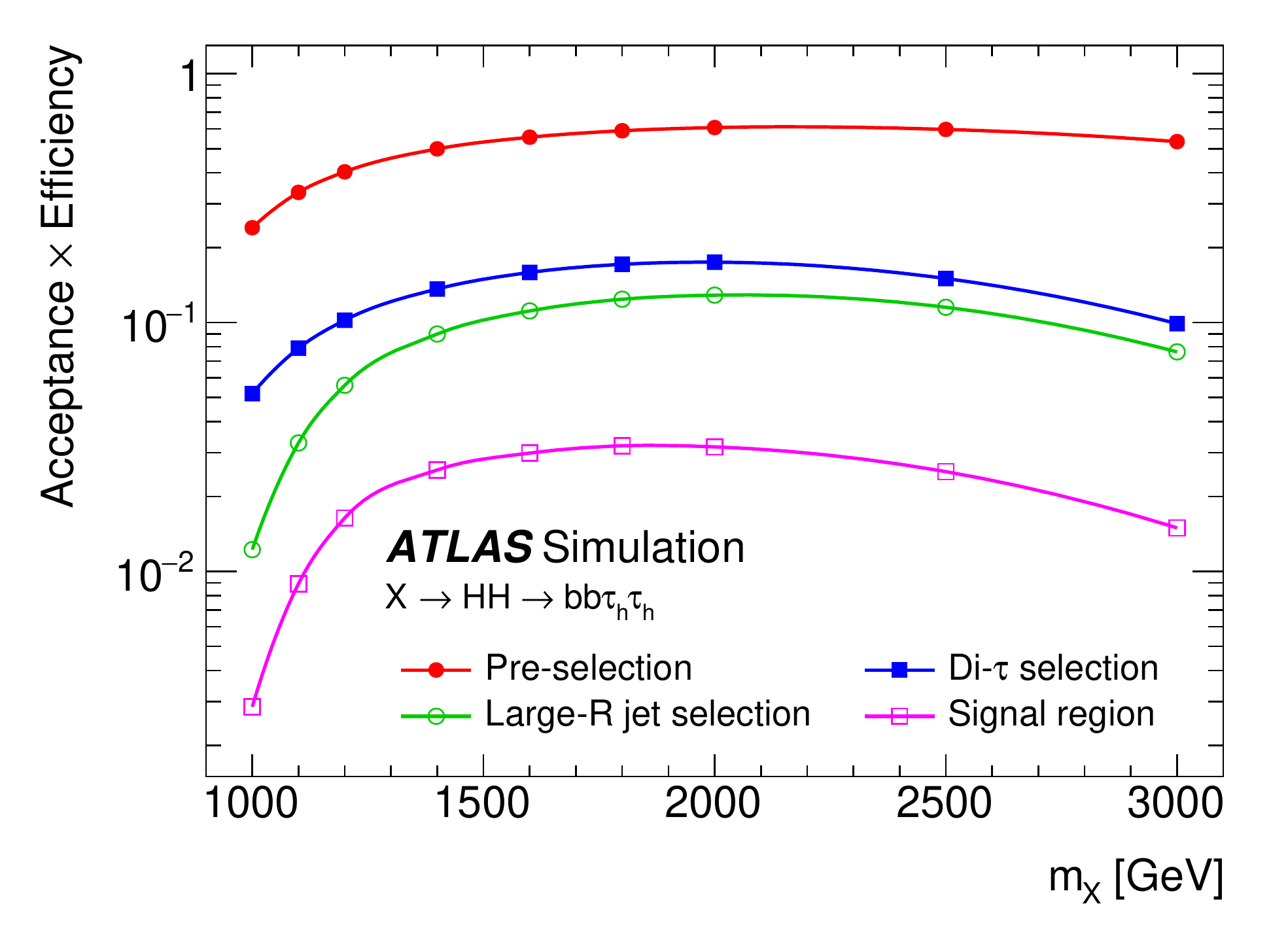}
\end{center}
\caption{Signal acceptance times selection efficiency as a function of the resonance mass, at various stages of
the event selection. From top to bottom: an event preselection (red solid-circle: trigger, object definitions and
$\met > 10\,\GeV$) is performed first; the requirements on the di-$\tau$ object and large-$R$ jet detailed in
the text are then applied (blue solid-box and green open-circle, respectively); finally, the $HH$ SR definition
must be satisfied (pink open-box).}
\label{fig:fig_eff_acc_1}
\end{figure}
 
In the SR, the dominant backgrounds arise from multi-jet, $ZH$, $Z$+hf and $Z$+lf events. The
multi-jet background with a misidentified di-$\tau$ object is estimated with the data-driven method
of Section~\ref{sec:ddbkgest}. The distribution shapes of events from the other background processes,
with a true di-$\tau$ object, are modelled using simulation, but the data-driven SF computed in
Section~\ref{sec:ditau-calib} is employed for their normalisation.
In addition, the \sherpa event generator was found to underestimate the
heavy-flavour contribution in the $Z$+jets process in previous ATLAS analyses, such as in
Ref.~\cite{HIGG-2018-04}. Therefore, the normalisation of the $Z \to \tau^+\tau^-$+hf background
is further corrected by a normalisation factor derived in a dedicated region enriched with
$Z \to \ell^+\ell^-$+hf events, where $\ell$ denotes either an electron or a muon defined as a `loose' lepton
in Ref.~\cite{HIGG-2018-04}. This region is selected by requiring two isolated electrons or muons with an invariant
mass $81\,\GeV < m_{\ell\ell} < 101\,\GeV$ and a dilepton transverse momentum $\pt^{\ell\ell} > 300\,\GeV$.
In addition, the selected events must contain at least one large-$R$ jet with
$\pt^{J} > 250\,\GeV$, $m_{J} > 50\,\GeV$ and two associated $b$-tagged track-jets.
After applying these event selections, the fraction of $Z$+hf events is about 80\%.
Figure~\ref{fig:fig_zllcr} shows a comparison of the predicted and measured $\pt^{\ell\ell}$
distributions. After subtracting all other processes from the data and comparing with simulated $Z$+hf
events, the normalisation factor is found to be $1.20 \pm 0.10\text{ (stat)} \pm 0.28\text{ (syst)}$.
The statistical uncertainty accounts for the limited size of the data and simulated event samples,
and is respectively 8.3\% and 1.7\% relative to the normalisation factor. Systematic uncertainties arise from
the instrumental sources detailed in Section~\ref{sec:ditau-calib} (3.4\%) and from neglecting the differences
between the \pT\ distributions of the visible decay products of $Z \to \ell^+\ell^-$ and $Z \to \tau^+\tau^-$, due to
the presence of neutrinos in the latter case (23\%).
 
\begin{figure}[htbp]
\begin{center}
\includegraphics[width=0.5\textwidth]{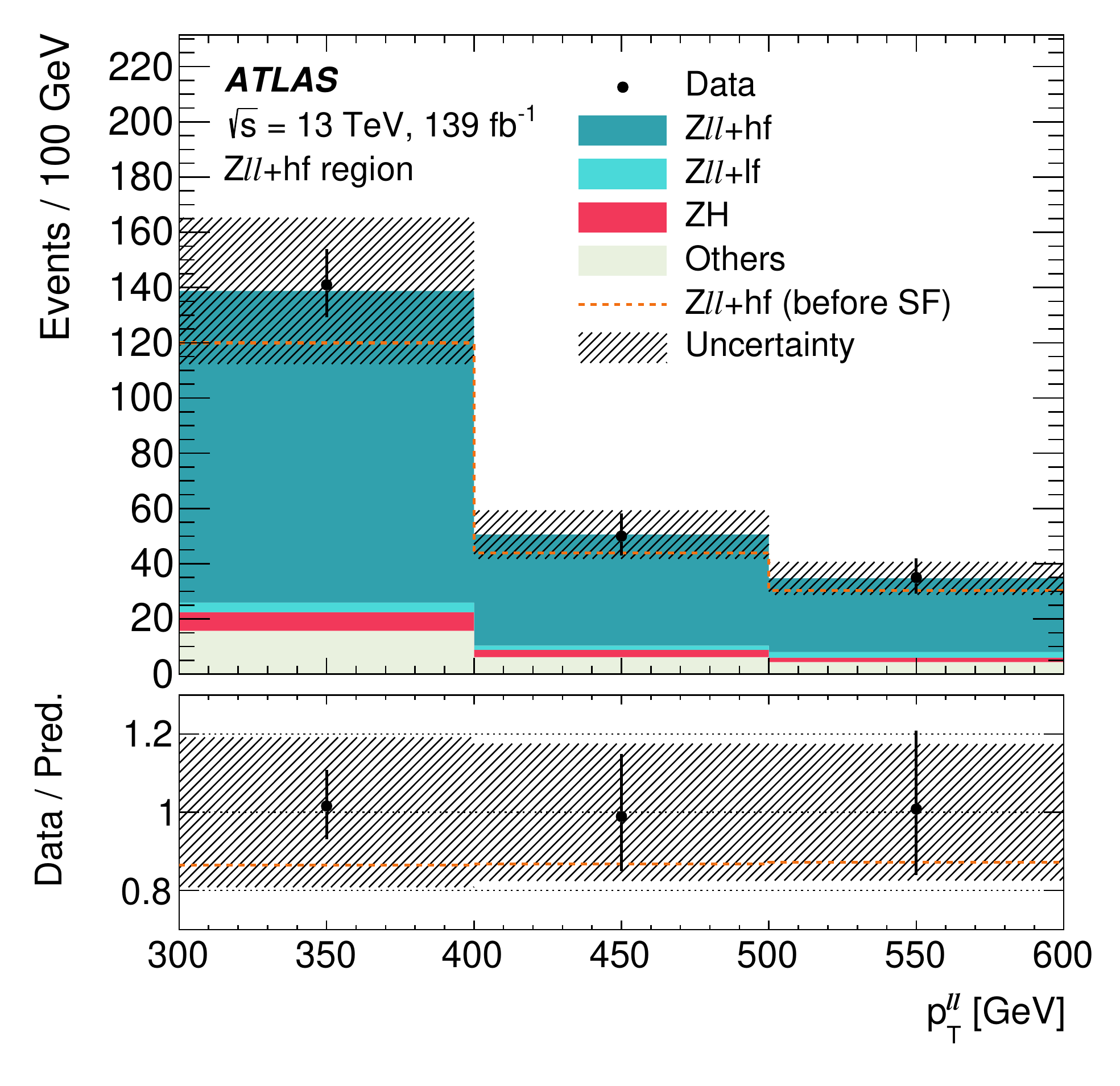}
\end{center}
\caption{Distribution of $\pt^{\ell\ell}$ in $Z \to \ell^+\ell^-$ events produced in association with
heavy-flavour jets. The background labelled as `Others' contains $W$+jets, diboson, \ttbar\ and
single-top-quark events.
The lower panel shows a bin-by-bin comparison of the data and the predicted total background,
in terms of event yield ratios.
The hatched bands represent combined statistical and systematic uncertainties.}
\label{fig:fig_zllcr}
\end{figure}
 
Table~\ref{tab:yields} summarises the event yields for data and all backgrounds, together with
the associated statistical and systematic uncertainties. The systematic uncertainties in the
multi-jet background estimation and of instrumental origin are the same as those discussed in
Sections~\ref{sec:ddbkgest} and~\ref{sec:ditau-calib}. The uncertainty related to the cross-sections
and acceptances of the simulated background processes is found to be negligible compared to the statistical
uncertainty. None of the simulated \ttbar\ events pass the event selection of the $HH$ SR, and hence
a $+1\sigma$ uncertainty of 0.12 events in the associated yield is estimated by considering the simulated events that
survive the selection criteria prior to applying requirements on $m_J$ and $m_{HH}^{\text{vis}}$.
 
The systematic uncertainties in the modelling of the signal acceptance and efficiency arise from four sources.
Those due to missing higher-order corrections are estimated by changing the renormalisation and factorisation
scales up and down by a factor of two. Systematic uncertainties arising from the strong coupling parameter are
assessed by varying its value. Uncertainties related to the PDF set used for matrix-element generation are
computed by comparing the signal acceptance and efficiency across the 100 replicas of the nominal NNPDF2.3LO PDF
set, as well as with alternative PDF sets, MMHT2014LO~\cite{Harland-Lang:2014zoa} and CT14~\cite{Dulat:2015mca}.
Finally, the impact of the parton shower and hadronisation modelling is assessed by comparing the signal
acceptance and efficiency between \textsc{Herwig~7} and \textsc{Pythia~8} predictions. The latter is the dominant
systematic uncertainty in the signal modelling (2--4\% for most of the mass hypotheses).
 
\begin{table}[htbp]
\centering
\caption{Event yields of the various estimated backgrounds and data, computed in the SR
of the search for $X \to HH \to b\bar{b}\tau^+\tau^-$. The background labelled as `Others' contains
$W$+jets, diboson, \ttbar\ and single-top-quark events. Statistical and systematic uncertainties are
quoted, in that order. The background yields and uncertainties are pre-fit and are found to be similar
to those post-fit. A dash '-' indicates a yield or uncertainty below 0.01.}
\label{tab:yields}
\small
\renewcommand{\arraystretch}{1.4}
\sisetup{retain-explicit-plus}
\begin{tabular}{cccc}
\toprule
Selection on $m^{\text{vis}}_{HH}$ & $>0$\,\GeV\ & $>900$\,\GeV\ & $>1200$\,\GeV \\
\midrule
$Z\tau\tau$+hf      & $ 0.89 \pm 0.25 ^{\num{+0.37}}_{\num{-0.35}}~~~~ $
& $ 0.75 \pm 0.21 ^{\num{+0.47}}_{\num{-0.37}}~~~~ $
& $ 0.17 \pm 0.05 \pm 0.07 $
\\
$Z\tau\tau$+lf      & $ 0.05 \pm 0.05 \pm 0.03 $
& $ 0.05 \pm 0.05 \pm 0.03 $
& -   \\
Multi-jet           & $ 0.18 \pm 0.03 \pm 0.14 $
& $ 0.17 \pm 0.03 \pm 0.13 $
& $ 0.09 \pm 0.02 \pm 0.07 $
\\
$ZH$                & $ 0.11 \pm 0.01 \pm 0.04 $
& $ 0.09 \pm 0.01 \pm 0.03 $
& $ 0.02 \pm \,\,\,~\text{-}~\,\,\, \pm 0.01 $
\\
Others              & $ 0.13 \pm 0.05 ^{\num{+0.15}}_{\num{-0.07}}~~~~ $
& $ 0.13 \pm 0.05 ^{\num{+0.15}}_{\num{-0.07}}~~~~ $
& $ 0.05 \pm 0.03 ^{\num{+0.12}}_{\num{-0.03}}~~~~ $
\\ \midrule
Sum of backgrounds         & $ 1.36 \pm 0.26 ^{\num{+0.42}}_{\num{-0.38}}~~~~ $
& $ 1.19 \pm 0.23 ^{\num{+0.51}}_{\num{-0.40}}~~~~ $
& $ 0.33 \pm 0.07 ^{\num{+0.16}}_{\num{-0.10}}~~~~ $
\\ \midrule
Data            & 2 & 2 & 0  \\ \hline
\bottomrule
\end{tabular}
\end{table}
 
Figure~\ref{fig:fig_mhh_2tag_sr} shows the $m^{\text{vis}}_{HH}$ distribution in the SR, prior
to the final selection using that variable. No significant excess with respect to the background expectation
is found and only two events are observed, with $m^{\text{vis}}_{HH}$ values of 1012 and 1114\,\GeV.
 
\begin{figure}[htbp]
\begin{center}
\includegraphics[width=0.5\textwidth]{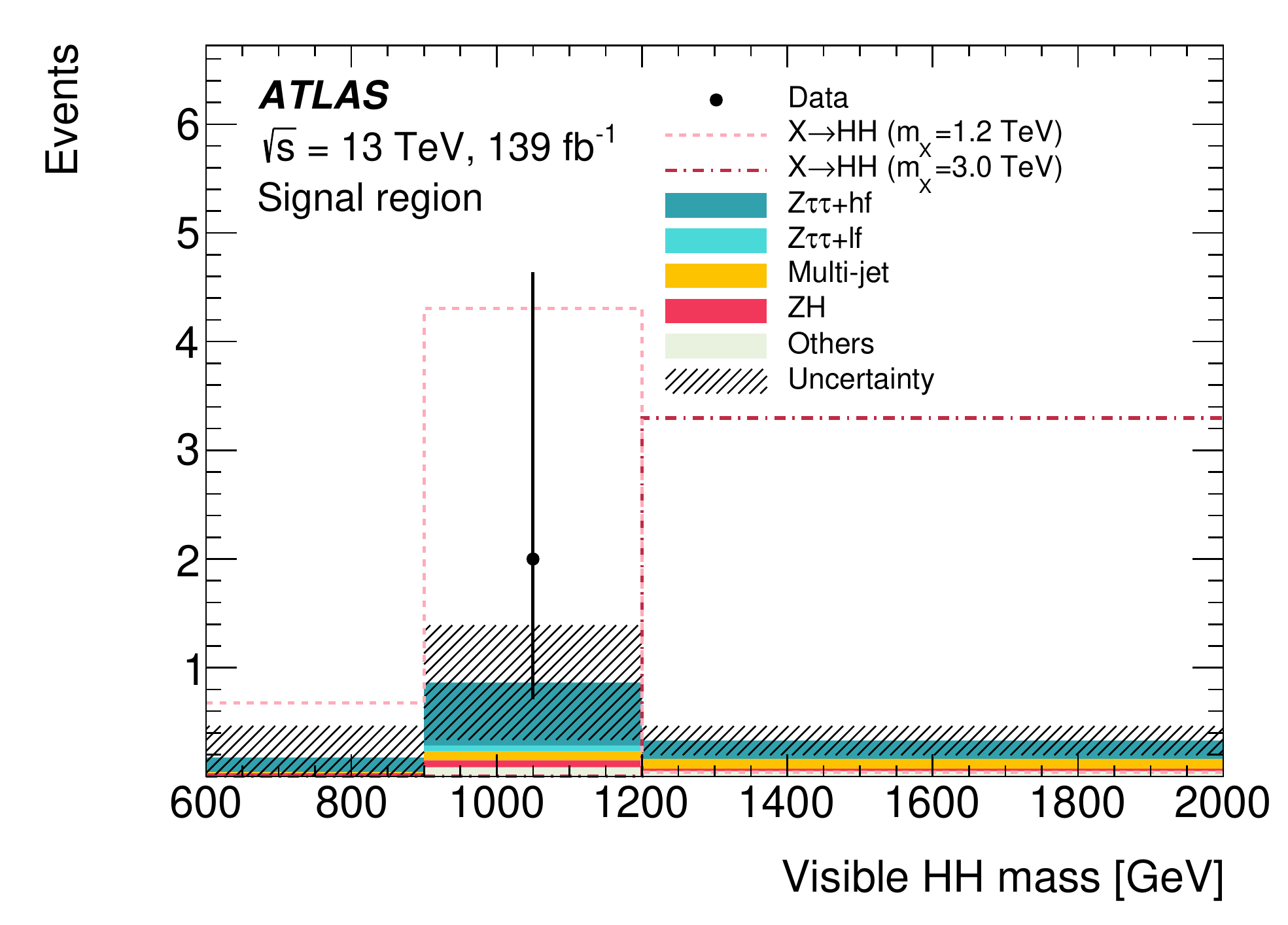}
\end{center}
\caption{Distributions of $m^{\text{vis}}_{HH}$ after applying all the event selections that define the
$HH$ SR, except the requirement on $m^{\text{vis}}_{HH}$. The background labelled as `Others'
contains $W$+jets, diboson, \ttbar and single-top-quark events. The $X \to HH \to b\bar{b}\tau^+\tau^-$
signal is overlaid for two resonance mass hypotheses with a cross-section set to the expected limit, while
all backgrounds are pre-fit. The first and the last bins contain the underflow and overflow bin entries,
respectively. The hatched bands represent combined statistical and systematic uncertainties.}
\label{fig:fig_mhh_2tag_sr}
\end{figure}
 
In order to extract exclusion limits on the $X \to HH$ signal cross-section, a single-bin counting experiment
is performed for every mass hypothesis, based on a profile-likelihood fit~\cite{Cowan:2010js}. All sources of
systematic uncertainty are accounted for as nuisance parameters that are profiled in the fit.
The 95\% CL upper limits on the $X \to HH$ cross-section are calculated as a function of $m_X$ using the CL$_\text{s}$
method~\cite{Read:2002hq}. Because both the expected and observed event yields are small, the distribution
of the test statistic is sampled by generating pseudo-experiments. For each mass hypothesis, 10\,000
background-only and 10\,000 signal+background pseudo-datasets are produced, and the exclusion limits are
obtained by scanning the signal strength and interpolating to find the value at which CL$_\text{s} = 0.05$. The
corresponding upper limits on the production cross-section of a heavy, narrow, scalar resonance decaying
into a pair of Higgs bosons are shown in Figure~\ref{fig:fig_xs_limits}.
 
Two different requirements are applied to $m^{\text{vis}}_{HH}$ for the resonance mass hypotheses of 1.6\,\TeV\
and 2.5\,\TeV, leading to discontinuities in the limits (at 1.6\,\TeV, the difference between imposing no
requirement and $m^{\text{vis}}_{HH} > 900\,\GeV$ is less than 1\% though). The more stringent requirement
on $m^{\text{vis}}_{HH}$ reduces the expected background for resonance mass hypotheses beyond 2.5\,\TeV,
and no data events survive. This leads to an observed exclusion limit slightly below the expected limit.
For resonance masses in the range 1.2--3.0\,\TeV, the observed (expected) 95\% CL upper limits lie between
94 and 28~fb (74 and 32~fb). For $m_X = 1.0\,\TeV$, where the analysis is least optimal, mostly due to the
\pt\ requirements on the large-$R$ jet at both the trigger and analysis levels, the observed (expected)
limit becomes 817~fb (624~fb). These 95\% CL upper limits are driven mainly by the statistical uncertainties
of the data sample. The relative impact of the systematic uncertainties is estimated by comparing the
expected upper limits when taking or not into account the nuisance parameters in the statistical analysis.
For resonance mass hypotheses in the range 1.2--2.0\,\TeV, the expected upper limits deteriorate by 20--30\%,
mostly due to the uncertainties in the di-$\tau$ reconstruction. At lower masses, the uncertainties in the
large-$R$ jet energy and mass scales further affect the expected upper limits (the largest impact from all
nuisance parameters reaches 85\% at 1\,\TeV). On the other hand, systematic uncertainties are found to have
a negligible impact (less than 10\%) on the expected upper limits for mass hypotheses beyond 2.5\,\TeV.
 
\begin{figure}[htbp]
\begin{center}
\includegraphics[width=0.55\textwidth]{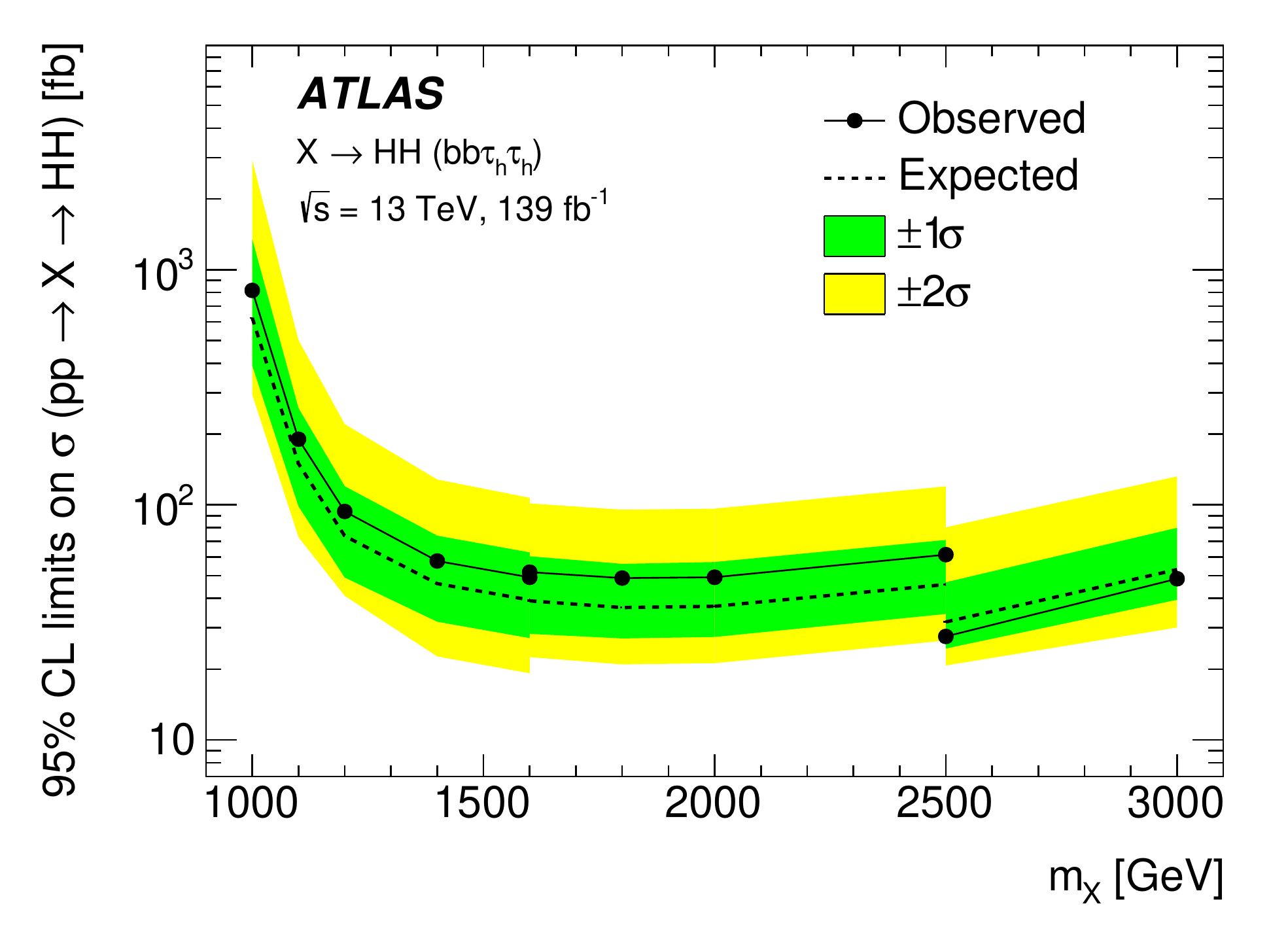}
\end{center}
\caption{Expected and observed 95\% CL upper limits on the production cross-section of a heavy, narrow,
scalar resonance decaying into a pair of Higgs bosons ($X \to HH$). The final state used in the search consists of
a boosted $b\bar{b}$ pair and a boosted hadronically decaying $\tau^+\tau^-$ pair, and the SM branching fractions
of the Higgs boson are assumed. The $\pm 1\sigma$ and $\pm 2\sigma$ variations about the expected limit are
indicated by the green and yellow bands, respectively.
}
\label{fig:fig_xs_limits}
\end{figure}


\FloatBarrier
 
\section{Conclusion}
\label{sec:conclusion}
In this paper, analysis techniques using large-radius jets and exploiting jet substructure are used
in the ATLAS experiment to reconstruct and identify boosted di-$\tau$ systems. These can arise from
the decay $H \to \tau^+\tau^-$ of a Higgs boson with a large Lorentz boost. A dedicated benchmark of the
di-$\tau$ tagger is designed to search for resonant Higgs boson pair production in the high-mass regime,
$X \to HH \to b\bar{b}\tau^+\tau^-$, where only final states with two hadronically decaying $\tau$-leptons are
considered. This benchmark corresponds to a 60\% di-$\tau$ identification efficiency and a rejection factor of $10^4$
for the background of quark- or gluon-initiated jets misidentified as boosted di-$\tau$ objects. The efficiency
of the di-$\tau$ tagger is determined using 139\,\ifb\ of LHC proton--proton collision data at $\sqrt{s}= 13\,\TeV$.
 
The search for resonant Higgs boson pair production via gluon--gluon fusion is carried out using the same dataset.
No significant deviations from the Standard Model predictions are observed, and 95\% confidence-level upper limits
are set on the production cross-section of a heavy, narrow, scalar resonance decaying into a pair of Higgs
bosons. Assuming SM branching fractions for the Higgs boson, the observed (expected) limits on the $X \to HH$
production cross-section lie between 94 and 28~fb (74 and 32~fb) for resonance mass hypotheses in the range
1.2--3.0\,\TeV.

\section*{Acknowledgements}
 

We thank CERN for the very successful operation of the LHC, as well as the
support staff from our institutions without whom ATLAS could not be
operated efficiently.
 
We acknowledge the support of ANPCyT, Argentina; YerPhI, Armenia; ARC, Australia; BMWFW and FWF, Austria; ANAS, Azerbaijan; SSTC, Belarus; CNPq and FAPESP, Brazil; NSERC, NRC and CFI, Canada; CERN; ANID, Chile; CAS, MOST and NSFC, China; COLCIENCIAS, Colombia; MSMT CR, MPO CR and VSC CR, Czech Republic; DNRF and DNSRC, Denmark; IN2P3-CNRS and CEA-DRF/IRFU, France; SRNSFG, Georgia; BMBF, HGF and MPG, Germany; GSRT, Greece; RGC and Hong Kong SAR, China; ISF and Benoziyo Center, Israel; INFN, Italy; MEXT and JSPS, Japan; CNRST, Morocco; NWO, Netherlands; RCN, Norway; MNiSW and NCN, Poland; FCT, Portugal; MNE/IFA, Romania; MES of Russia and NRC KI, Russia Federation; JINR; MESTD, Serbia; MSSR, Slovakia; ARRS and MIZ\v{S}, Slovenia; DST/NRF, South Africa; MICINN, Spain; SRC and Wallenberg Foundation, Sweden; SERI, SNSF and Cantons of Bern and Geneva, Switzerland; MOST, Taiwan; TAEK, Turkey; STFC, United Kingdom; DOE and NSF, United States of America. In addition, individual groups and members have received support from BCKDF, CANARIE, Compute Canada and CRC, Canada; ERC, ERDF, Horizon 2020, Marie Sk{\l}odowska-Curie Actions and COST, European Union; Investissements d'Avenir Labex, Investissements d'Avenir Idex and ANR, France; DFG and AvH Foundation, Germany; Herakleitos, Thales and Aristeia programmes co-financed by EU-ESF and the Greek NSRF, Greece; BSF-NSF and GIF, Israel; La Caixa Banking Foundation, CERCA Programme Generalitat de Catalunya and PROMETEO and GenT Programmes Generalitat Valenciana, Spain; G\"{o}ran Gustafssons Stiftelse, Sweden; The Royal Society and Leverhulme Trust, United Kingdom.
 
The crucial computing support from all WLCG partners is acknowledged gratefully, in particular from CERN, the ATLAS Tier-1 facilities at TRIUMF (Canada), NDGF (Denmark, Norway, Sweden), CC-IN2P3 (France), KIT/GridKA (Germany), INFN-CNAF (Italy), NL-T1 (Netherlands), PIC (Spain), ASGC (Taiwan), RAL (UK) and BNL (USA), the Tier-2 facilities worldwide and large non-WLCG resource providers. Major contributors of computing resources are listed in Ref.~\cite{ATL-SOFT-PUB-2020-001}.
 
 
\clearpage
 
\appendix

\printbibliography
 
\clearpage
 
\begin{flushleft}
\hypersetup{urlcolor=black}
{\Large The ATLAS Collaboration}

\bigskip

\AtlasOrcid[0000-0002-6665-4934]{G.~Aad}$^\textrm{\scriptsize 102}$,    
\AtlasOrcid[0000-0002-5888-2734]{B.~Abbott}$^\textrm{\scriptsize 128}$,    
\AtlasOrcid{D.C.~Abbott}$^\textrm{\scriptsize 103}$,    
\AtlasOrcid[0000-0002-2788-3822]{A.~Abed~Abud}$^\textrm{\scriptsize 36}$,    
\AtlasOrcid[0000-0002-1002-1652]{K.~Abeling}$^\textrm{\scriptsize 53}$,    
\AtlasOrcid[0000-0002-2987-4006]{D.K.~Abhayasinghe}$^\textrm{\scriptsize 94}$,    
\AtlasOrcid[0000-0002-8496-9294]{S.H.~Abidi}$^\textrm{\scriptsize 166}$,    
\AtlasOrcid[0000-0002-8279-9324]{O.S.~AbouZeid}$^\textrm{\scriptsize 40}$,    
\AtlasOrcid{N.L.~Abraham}$^\textrm{\scriptsize 155}$,    
\AtlasOrcid[0000-0001-5329-6640]{H.~Abramowicz}$^\textrm{\scriptsize 160}$
% ...
\bigskip
\\

% ...
$^{36}$CERN, Geneva; Switzerland.\\
% ...
$^{40}$Niels Bohr Institute, University of Copenhagen, Copenhagen; Denmark.\\
% ...
$^{53}$II. Physikalisches Institut, Georg-August-Universit\"{a}t G\"ottingen, G\"ottingen; Germany.\\
% ...
$^{94}$Department of Physics, Royal Holloway University of London, Egham; United Kingdom.\\
% ...
$^{102}$CPPM, Aix-Marseille Universit\'e, CNRS/IN2P3, Marseille; France.\\
$^{103}$Department of Physics, University of Massachusetts, Amherst MA; United States of America.\\
% ...
$^{128}$Homer L. Dodge Department of Physics and Astronomy, University of Oklahoma, Norman OK; United States of America.\\
% ...
$^{155}$Department of Physics and Astronomy, University of Sussex, Brighton; United Kingdom.\\
% ...
$^{160}$Raymond and Beverly Sackler School of Physics and Astronomy, Tel Aviv University, Tel Aviv; Israel.\\
% ...
$^{166}$Department of Physics, University of Toronto, Toronto ON; Canada.\\
% ...

\end{flushleft}


\newpage

\end{document}